\documentclass[aps,prd,11pt,preprint,nofootinbib,showkeys,superscriptaddress]{revtex4-1} 
\usepackage{amsmath,amssymb,amsthm,mathrsfs,bbm}
\usepackage{latexsym,amscd,amsbsy,amsfonts,dsfont}
\usepackage{graphicx}
\usepackage[utf8]{inputenc}
\usepackage{marvosym}
\usepackage{float}
\usepackage{slashed}
\usepackage{cancel}
\usepackage{multirow}
\usepackage{soul}
\usepackage{physics}
\usepackage{comment}
\usepackage[normalem]{ulem}
\usepackage[usenames,dvipsnames]{xcolor}
\usepackage[
      colorlinks=true,
      linktocpage=true,
      breaklinks=true,
      linkcolor=Aquamarine,
      urlcolor=Purple,
      filecolor=black,
      citecolor=Purple,
      menucolor=black,
      pdfstartview=FitV,
      bookmarksopen=true
      ]{hyperref}
\usepackage{tikz}
\usepackage[usenames,dvipsnames]{xcolor}
\graphicspath{{Figures/},{Logos/}}

\begin{document}

\title{Duality Symmetry and Anomaly for Gravitational Waves in Curved Spacetimes}

\author{Adrian del Rio}
\email{adrdelri@math.uc3m.es}
\affiliation{Universidad Carlos III de Madrid, Departamento de Matem\'aticas.\\ Avenida de la Universidad 30 (edificio Sabatini), 28911 Legan\'es (Madrid), Spain.}

\author{Javier Olmedo}
\email{javolmedo@ugr.es}
\affiliation{Departamento de F\'isica Te\'orica y del Cosmos, Universidad de Granada, Granada-18071, Spain}
\author{Ant\'onio Torres Manso}
\email{atmanso@uc.pt}
\affiliation{Universidad de Coimbra, Faculdade de Ci\^encias e Tecnologia da Universidade de Coimbra and CFisUC, Rua Larga, 3004-516 Coimbra, Portugal}

\begin{abstract}

The vacuum Einstein equations admit a formulation closely analogous to the source-free Maxwell theory. In particular, the linearized equations exhibit an electric-magnetic duality symmetry. We develop a framework that makes this analogy manifest by explicitly identifying the electric and magnetic components of perturbative gravitational waves. Within this formulation, we show that duality rotations between these gravitoelectric and gravitomagnetic fields constitute a Noether symmetry of the linearized theory, and we derive the associated conserved current. The corresponding conserved charge encodes the difference in intensity between the right- and left-handed circularly polarized components of the gravitational wave — that is, between its self-dual and anti-self-dual parts. Remarkably, this conservation law remains valid even when the gravitational perturbations propagate on generic curved backgrounds. We then investigate whether this symmetry survives quantization. While the duality symmetry is preserved at the quantum level in flat spacetime, we find that it is anomalously broken in curved backgrounds. As a result, an imbalance between right- and left-handed gravitons could be excited from the vacuum. This effect represents a chiral anomaly for massless spin-two fields, generalizing known results for fermions and spin-one photon fields.


\end{abstract}

\maketitle

\section{Introduction}
\label{Sec:Introduction}

Symmetries have always played a fundamental role in theoretical physics, serving as guiding principles in the construction of physical theories. Conservation laws, as described by Noether’s theorem, have been essential to our understanding of both particle physics and gravity. Among many examples, one that often flies under the radar is the electric-magnetic duality symmetry in free Maxwell theory\footnote{ The symmetry has also been extended to the non-linear regime in \cite{Avetisyan:2021heg}.}. In its continuous $U(1)$ version, this symmetry has recently been linked to a chiral symmetry, a concept traditionally associated with fermionic fields \cite{Agullo:2018nfv}.

However, it is well known that classical Noether symmetries may not survive quantization, giving rise to so-called quantum anomalies \cite{alma991005794549706258}. These were first uncovered by Adler, Bell, and Jackiw in the context of the pion decay puzzle \cite{Adler:1969gk,Bell:1969ts}. They showed that the classical chiral symmetry of a massless Dirac field is broken at the quantum level when coupled to an electromagnetic background. A similar anomaly arises when a massless Dirac field interacts with a classical gravitational background \cite{Kimura:1969iwz,Delbourgo:1972xb,Eguchi:1976db}.
Interestingly, this quantum effect is closely connected to particle production in strong backgrounds—both gravitational and electromagnetic \cite{PhysRevD.100.085014, Domcke:2019qmm,Navarro-Salas:2021sex,Navarro-Salas:2022zeq, PhysRevD.108.105025}. A prototypical case that combines both is the Schwinger effect in de Sitter space \cite{Bastero-Gil:2025jio}. For recent advances in particle production from electromagnetic fields, see also \cite{Fedotov:2022ely,Pla:2020tpq,Alvarez-Dominguez:2023ten} and references therein.

More recently, it has been shown that electric-magnetic duality transformations are also anomalous provided the electromagnetic fields propagate in a nontrivial gravitational background \cite{Agullo:2016lkj,Agullo:2017pyg,Agullo:2018iya,Agullo:2018nfv,delRio:2020cmv,PhysRevD.104.065012, PhysRevD.108.044052}. To arrive at this result, Maxwell’s theory was reformulated in terms of self-dual and anti-self-dual variables. In this framework, duality transformations become formally analogous to chiral transformations of massless spin-$1/2$ Dirac fields, leading to a spin-1 generalization of the chiral anomaly. This naturally raises the question: What happens to gravitational waves, i.e., massless spin-2 fields, in such backgrounds?


Duality symmetry is also a feature of the nonlinear vacuum Einstein equations \cite{Ellis:1973,RoyMaartens1998,Dadhich:1999eh,Kol:2022bsd}, where the Weyl tensor can be decomposed into electric and magnetic (symmetric, traceless) parts. The Bianchi identities provide evolution equations for these components, yielding a structure reminiscent of Maxwell’s theory. However, the full quantization of gravity is notoriously difficult, and thus, to answer the question above, we turn our attention to linearized gravity.\footnote{A related construction in asymptotically flat spacetimes has been studied in \cite{Seraj:2021rxd, Seraj:2022qyt}.}

The formulation of duality transformations—or self-dual descriptions—in linearized gravity has been explored for some time \cite{Henneaux:2004jw,Deser:2005sz,Aghapour:2018pup,Aghapour:2021bkb,Toth:2021dut}. In particular, it has been shown that, within Einstein's theory, gravitational perturbations in flat spacetime (gravitational waves) obey equations of motion that can be cast in a form analogous to Maxwell’s equations \cite{Aghapour:2018pup,Aghapour:2021bkb}. Within this framework, duality transformations correspond to electric-magnetic rotations of the field components.
In this manuscript, we build on these ideas. Inspired by the suggestion in \cite{Agullo:2018nfv}, we reformulate the system using self-dual and anti-self-dual variables, and recast the dynamics into a Dirac-like equation. This reformulation allows us to study the helicity structure of gravitational waves in terms of  chirality of some spinors. We then extend the formalism to curved backgrounds using covariant techniques, while keeping the analysis within the geometric optics approximation.

In this paper, we demonstrate that quantum fluctuations of gravitational perturbations can break the classical duality symmetry when the gravitational waves propagate in a curved background\footnote{Discrete $\mathbb Z_2$   duality transformations have also  been studied in the context of string theories. While  not related to continuous Noether symmetries and associated quantum anomalies, they are also regarded ``anomalous'' in the sense that partition functions fail to be modular-invariant. This was first studied in electromagnetism \cite{Witten:1995gf, PhysRevD.96.045008} and recently in linearized gravity \cite{Borsten:2025phf}.}. We derive this result by computing the vacuum expectation value of the classically conserved current, using heat kernel renormalization techniques.


The paper is organized as follows. In Sec. \ref{sectionflatGWs}, we introduce the electric and magnetic components of perturbative gravitational waves in flat spacetime. Section \ref{sectionselfdual} presents the formulation in terms of self-dual and anti-self-dual variables and connects it to a Dirac-like description. In Sec. \ref{sec:curvedbd}, we generalize the discussion to arbitrary curved spacetimes. The anomaly in the duality symmetry is addressed in Sec. \ref{sec:anomaly}, and our conclusions are presented in Sec. \ref{sec:concl}. Five appendices are included to make the manuscript self-contained.


Throughout this work, we use Greek indices $\mu, \nu,\ldots$ for tensors in curved spacetimes with metric signature $(-,+,+,+)$; Latin indices $a, b, c,\ldots$ for tensors in Minkowski spacetime; and $i, j, k,\ldots$ for spatial indices. Internal indices $I, J, K,\ldots$ or $\dot{I}, \dot{J}, \dot{K},\ldots$ refer to spin-1 representations of the complex Lorentz group and their metric signature is $(+,-,-,-)$. 

\section{Duality symmetry  for gravitational perturbations on a flat background: A path to a Maxwell equation}\label{sectionflatGWs}


In this section, we will analyze linearized gravitational waves propagating on a flat spacetime and show that they exhibit a structure remarkably similar to source-free Maxwell theory. By adopting the transverse-traceless gauge and introducing suitable analogues of electric and magnetic fields for gravity, we will recast the dynamics of linearized gravity into a Maxwell-like form. This formulation makes the duality symmetry of the system manifest and will allow us to identify a conserved current associated with helicity. 

We begin with the Einstein-Hilbert Lagrangian,
\begin{equation}
	\mathcal{L}_{\mathrm{EH}}=\frac{\sqrt{-g}}{16 \pi G} R,
\end{equation}
where $g_{\mu\nu}$ is the spacetime metric and $R$ the Ricci scalar. We consider linear perturbations around flat Minkowski space,  $g_{\mu\nu}=\eta_{\mu\nu}+h_{\mu\nu}$, where $\eta_{\mu\nu}$ is the flat metric and $h_{\mu\nu}$ denotes the perturbation. Plugging this expansion into the equation above, integrating by parts to remove second derivatives (discarding total derivatives in the process),{  and neglecting higher-order terms than quadratic in the perturbative expansion,} we obtain the linearized gravity Lagrangian—known as the Fierz–Pauli Lagrangian \cite{Fierz:1939ix,deRham:2014zqa}.
\begin{equation}
	\mathcal{L}_{\mathrm{LG}}=\frac{1}{2}\left(\partial_b h^a{ }_a \partial^b h^c{ }_c-2 \partial_b h^a{ }_a \partial^c h^b{ }_c-\partial_c h_{a b} \partial^c h^{a b}+2 \partial_c h_{a b} \partial^b h^{a c}\right),
\end{equation}
where we have set $32\pi G=1$. Here, $\partial_a$ denotes partial derivatives in a global inertial coordinate system of Minkowski spacetime. We use the notation $h_{ab}$ with Latin indices to emphasize that it is a symmetric tensor defined on flat spacetime, and all indices are raised and lowered with the background Minkowski metric $\eta_{ab}$.  Imposing the transverse-traceless (TT) gauge 
\begin{equation}
	h_{0 a}=0, \quad h_{a b}{ }^{,b}=0, \quad h^a{ }_a=0,\label{Eq:TTgauge}
\end{equation}
only the spatial components of the perturbation, denoted by $h_{ij}$, are nontrivial. The Fierz–Pauli Lagrangian then simplifies to
\begin{equation}\label{eq:LG2}
	\mathcal{L}_{\mathrm{LG}}=\frac{1}{2}\left(\dot{h}_{i j} \dot{h}^{i j}-h_{i j, k} h^{i j, k}+2 h_{i j, k} h^{i k, j}\right) .
\end{equation}
Adding now a total derivative yields an equivalent Lagrangian:
\begin{equation}
	\begin{aligned}
		\mathcal{L}_{\mathrm{LG}}^{\prime} & =\mathcal{L}_{\mathrm{LG}}-\frac{1}{2}\left(h_{j k} h^{a k, j}\right)_{, a}\\
	 & =\mathcal{L}_{\mathrm{LG}}-\frac{1}{2}\left(h_{j k} h^{i k, j}\right)_{, i} \\ \label{eq:LG3}
		& =\frac{1}{2}\left(\dot{h}_{i j} \dot{h}^{i j}-h_{i j, k} h^{i j, k}+h_{i j, k} h^{i k, j}\right).
	\end{aligned}
\end{equation}
where the second equality follows from  $\left(h_{j k} h^{0 k, j}\right)_{, 0}=0$, valid in the TT gauge.

Following  \cite{Aghapour:2018pup}, we now introduce specific notation  that makes tensor manipulations resemble familiar vector operations. Let $\textit{\textbf{c}}$ and $\textit{\textbf{d}}$ denote symmetric, traceless rank-2 tensors (e.g., $c_{ij}$). We define the following binary operations:
\begin{flalign}
	&\text{scalar dot product }\qquad \boldsymbol{c} \cdot \boldsymbol{d}=c_{i j} d^{i j},\\
	&\text{cross product} \qquad(\boldsymbol{c} \times \boldsymbol{d})_i=\epsilon_i^{\ j k} c_{j l} d_k^{\,\ l},\\
	&\text{2-tensor dot product} \qquad(\boldsymbol{c}: \boldsymbol{d})_{i j}=c_{k(i} d_{j)}{ }^k,\\
	&\text{wedge product} \qquad(\boldsymbol{c} \wedge \boldsymbol{d})_{i j}=\epsilon_i{ }^{k l} \epsilon_j{ }^{m n} c_{k m} d_{l n}\,,
\end{flalign}
where the curly bracket index notation denotes symmetrization.
We also define the divergence and curl of a symmetric tensor $\textbf{e}$ as:
\begin{equation}
	(\nabla \cdot \boldsymbol{e})_i=e_{i j}{ }^{,j}\,, \quad(\nabla \times \boldsymbol{e})_{i j}=\epsilon_{(i}{ }^{k l} e_{j) l, k}\, .
\end{equation}
Now we define the gravitoelectric and gravitomagnetic fields by the following symmetric, traceless, rank- 2 tensors: 
\begin{equation}\label{eq:enb}
	e_{i j}=-\dot{h}_{i j}, \quad b_{i j}=\epsilon_i^{\ l m} h_{j m, l}\,,
\end{equation}
or, in vector notation:
\begin{flalign}
	\boldsymbol{e}&=-\dot{\boldsymbol{h}}, 	\label{Eq:potentials&fields}\\
	\boldsymbol{b}&=\nabla \times \boldsymbol{h}.
\end{flalign}
These fields play the roles of the electric and magnetic fields in Maxwell theory, while ${h}_{i j}$ is the analogue of the vector potential. 
Furthermore, with this notation the Lagrangian becomes
\begin{equation}
	\begin{aligned}
		\mathcal{L}_{\mathrm{LG}}^{\prime} & =\frac{1}{2}[\dot{\boldsymbol{h}} \cdot \dot{\boldsymbol{h}}-(\nabla \times \boldsymbol{h}) \cdot(\nabla \times \boldsymbol{h})] \\
		& =\frac{1}{2}(\boldsymbol{e} \cdot \boldsymbol{e}-\boldsymbol{b} \cdot \boldsymbol{b}),\label{Eq:GWMaxwelllikeL}
	\end{aligned}
\end{equation}
which mirrors the structure of the Maxwell Lagrangian. The resulting Euler-Lagrange equations and Bianchi identities yield:
\begin{flalign}
	&\nabla \cdot \boldsymbol{e}=0, \quad \nabla \cdot \boldsymbol{b}=0, \\ &\nabla \times \boldsymbol{e}=-\dot{\boldsymbol{b}},\quad \nabla \times \boldsymbol{b}=\dot{\boldsymbol{e}},
\end{flalign} 
which, again, are very similar to Maxwell's equations. These equations are all symmetric under duality rotations of the electric and magnetic components. Specifically, given a solution to the above equations for $e_{i j}$ and $b_{i j}$, the combination 
\begin{flalign} \label{11}
	e_{i j} \rightarrow e_{i j} \cos \theta+b_{i j} \sin \theta,\\
	b_{i j} \rightarrow b_{i j} \cos \theta-e_{i j} \sin \theta, \label{22}
\end{flalign}
is also a solution. This is a continuous $U(1)$ symmetry analogous to electromagnetic duality in Maxwell theory, which in fact leaves  the Lagrangian  $\mathcal{L}_{\mathrm{LG}}^{\prime}$ invariant.

To describe this symmetry in terms of potentials, we introduce an auxiliary symmetric rank-2 tensor $k_{ij}$, satisfying the same TT gauge conditions as $h_{ij}$ in Eq. \eqref{Eq:TTgauge}, and analogous to the dual ``electric'' potential introduced for the study of the electromagnetic duality \cite{Aghapour:2018pup}. The two potentials are not independent from each other, but they obey the duality conditions:
%
\begin{equation}
	\begin{aligned}
		\dot{\boldsymbol{h}} & =\nabla \times \boldsymbol{k}, \\
		\dot{\boldsymbol{k}} & =-\nabla \times \boldsymbol{h}\,.
	\end{aligned}
\end{equation} 
The gravitoelectric and gravitomagnetic  fields $e_{ij}$ and $b_{ij}$ can then be written in terms of these potentials as 
\begin{flalign}
	\boldsymbol{e}&=-\dot{\boldsymbol{h}}=-\nabla \times \boldsymbol{k}, 	\label{Eq:potentials&fields}\\
	\boldsymbol{b}&=\nabla \times \boldsymbol{h}=-\dot{\boldsymbol{k}}\, ,
\end{flalign}
and duality rotations act as 
\begin{flalign}
	h_{i j} \rightarrow h_{i j} \cos \theta+k_{i j} \sin \theta,\\
	k_{i j} \rightarrow k_{i j} \cos \theta-h_{i j} \sin \theta .
\end{flalign}


The invariance 
under the continuous duality transformation leads to a  conservation law determined by the Noether current $J^a$, given by
\begin{flalign}
	J^0=&\frac{1}{2}\delta\theta\left(\dot{h}_{i j} k^{i j}-\dot{k}_{i j} h^{i j}\right)=\frac{1}{2}\delta\theta(\boldsymbol{h} \cdot \boldsymbol{b}-\boldsymbol{k} \cdot \boldsymbol{e}), \\
	J^i=&\frac{1}{2}\delta\theta \epsilon^{i j k}\left(e_{l j} h^l{ }_k+b_{l j} k_{\ k}^l\right)=\frac{1}{2}\delta\theta(\boldsymbol{e} \times \boldsymbol{h}+\boldsymbol{b} \times \boldsymbol{k})^i\,,
	\label{Eq:Currentmaxwellform}
\end{flalign}
and the analogous to the EM helicity current and conservation law
\begin{equation}
	\dot{\mathcal{H}}+\nabla \cdot \boldsymbol{S}=0 \,,
\end{equation} 
with helicity $\mathcal{H}=2 J^0$ and  spin $\boldsymbol{S}=2\boldsymbol{J}$.

\section{Self-dual variables and the spinorial formulation}\label{sectionselfdual}

The description of gravitational perturbations in terms of gravitoelectric and gravitomagnetic fields given in the previous section allowed us to highlight a structural analogy with Maxwell theory. In this section, we further reformulate this approach using self- and anti-self-dual variables, following \cite{Agullo:2018nfv}. This formulation provides a natural framework to describe the electric-magnetic duality as a chiral symmetry, making the analogy with fermions more transparent, and building a bridge towards an eventual spinorial formulation of the theory.

With this in mind, let us define the self- and anti-self-dual combinations of the linearized gravitational field as:
\begin{flalign}\label{eq:Hij}
	\mathcal{H}^\pm_{ij}=\frac{1}{\sqrt{2}}\left(e_{ij}\pm i\,b_{ij}\right),
\end{flalign}
where $e_{ij}$ and $b_{ij}$ are given by equations (\ref{eq:enb}), as in the previous section, and  the superscripts $\pm$ refer to the self-dual and anti-self-dual parts, respectively. Note that $\mathcal{H}_{ij}^+=\overline{\mathcal{H}_{ij}^-}$, and that both  are symmetric and traceless rank-2 tensors. These objects are the analogues of the self-dual and anti-self-dual parts of the Maxwell field $F_{\mu\nu}$, with respect to the Hodge dual operation. Furthermore, under  duality rotations \eqref{11}-\eqref{22} one gets
\begin{flalign}
	\mathcal{H}^\pm_{ij}\rightarrow e^{\mp i \theta}\mathcal{H}^\pm_{ij}\,,
\end{flalign}
and for $\theta=\pi/2$ we find $\mathcal{H}^\pm_{ij}\rightarrow \pm\, i \mathcal{H}^\pm_{ij}$, giving the meaning of self- and anti-self-dual denomination. 
In terms of these complex variables, the linearized Einstein's equations can now be expressed as
\begin{flalign}
	&\nabla \cdot \boldsymbol{\mathcal{H}}^\pm=0, \label{Eq:Maxwelleqselfdualdiv}\\ 
	&\nabla \times \boldsymbol{\mathcal{H}}^\pm=\pm i\dot{\boldsymbol{\mathcal{H}}}^\pm\,,\label{Eq:Maxwelleqselfdualrot}
\end{flalign}
which follow  directly from the Maxwell-like equations for \textit{$\textbf{e}$, $\textbf{b}$}. The dynamics of $\boldsymbol{\mathcal{H}}^\pm$ therefore decouples into two independent sectors, each associated with a definite duality (or chirality) component. The constraint in the first equation above can  be integrated to give the self-dual potentials
\begin{equation}
	\boldsymbol{\mathcal{H}}^\pm=\pm\, i\, \nabla \times \boldsymbol{{h}}^\pm\,.\label{Eq:potentialsdef}
\end{equation}
From this definition, we can write Maxwell's equations for the potentials by replacing Eq.~\eqref{Eq:potentialsdef} in the dynamical  equations in Eq.~\eqref{Eq:Maxwelleqselfdualrot}. Then, integrating the curl we find
\begin{flalign}
	\pm i\,\nabla \times \boldsymbol{h}^\pm=- \dot{\boldsymbol{h}}^\pm+\nabla \boldsymbol{h}_0^\pm\,,\label{Eq:Maxwelleqsselfdualpot}
\end{flalign}
where  $\nabla\boldsymbol{h}_0^\pm$ arises as a constant of integration. Here, $\boldsymbol{h}_0^\pm$ is  a vector that we identify as $h_{0i}$ analog to the potential $A_0$ in electromagnetism. Furthermore, $ \nabla\boldsymbol{h}_0^\pm=h^\pm_{0i,j}$ can be removed with the gauge choice in Eq.~\eqref{Eq:TTgauge}. 
Both Eq.~\eqref{Eq:Maxwelleqsselfdualpot} and Eqs.~\eqref{Eq:Maxwelleqselfdualdiv} and \eqref{Eq:Maxwelleqselfdualrot} describe the same dynamics for free linearized gravitational perturbations propagating in flat Minkowski space. 

As a side remark, let us note that, as in the electromagnetic case~\cite{Agullo:2018nfv}, we can write an analogue Gauss law, $\nabla \cdot \boldsymbol{e}=0$, to then define an auxiliary potential $\boldsymbol{k}$ as in Eq.~\eqref{Eq:potentials&fields}. Moreover, Maxwell's equations for the original potentials are written as 
\begin{flalign}
	&\nabla \times \boldsymbol{h}= \dot{\boldsymbol{k}}-\nabla \boldsymbol{k}_0\,,\nonumber\\
	&\nabla \times \boldsymbol{k}=- \dot{\boldsymbol{h}}+\nabla\boldsymbol{h}_0\,,
	\label{Eq:Maxwelleqsspot}
\end{flalign} where $\boldsymbol{h}_0$ and $\boldsymbol{k}_0$ are analogous to the scalar potentials. This allows us to recover our initial definitions for the self- and anti-self-dual potentials by ${\boldsymbol{h}^\pm=\frac{1}{\sqrt{2}}\left(\boldsymbol{h}\pm i\,\boldsymbol{k}\right)}$.

Introducing the matrices 
\begin{flalign}\nonumber
	\alpha_0^{a b}=\eta^{a b}=\left(\begin{array}{cccc}
		-1 & 0 & 0 & 0 \\
		0 & 1 & 0 & 0 \\
		0 & 0 & 1 & 0 \\
		0 & 0 & 0 & 1
	\end{array}\right), \qquad\alpha_1^{a b}=\left(\begin{array}{cccc}
		0 & -1 & 0 & 0 \\
		1 & 0 & 0 & 0 \\
		0 & 0 & 0 & i \\
		0 & 0 & -i & 0
	\end{array}\right), \\ 
	\alpha_2^{a b}=\left(\begin{array}{cccc}
		0 & 0 & -1 & 0 \\
		0 & 0 & 0 & -i \\
		1 & 0 & 0 & 0 \\
		0 & i & 0 & 0
	\end{array}\right), \qquad \alpha_3^{a b}=\left(\begin{array}{cccc}
		0 & 0 & 0 & -1 \\
		0 & 0 & i & 0 \\
		0 & -i & 0 & 0 \\
		1 & 0 & 0 & 0
	\end{array}\right),\label{eq:alphas}
\end{flalign}
one can rewrite Maxwell's equations for the potentials as 
\begin{equation}
	\bar{\alpha}_{(\dot I}^{a b}\, h_{c)b,a}^{+}=0\,,
	\qquad {\alpha}_{(I}^{a b}\, h_{c)b,a}^{-}=0\,,\label{Eq:EOMsdpotentials}
\end{equation}
where the bar over $\alpha^{ab}_I$ denotes complex conjugation. Indices $I, J, \ldots$ and $\dot I, \dot J, \ldots$ run from 0 to 3, just as the spacetime indices $a,b,\ldots$, and we have extended the potentials $\boldsymbol{h}$ to $4\times4$ matrices with $h_{0a}=\boldsymbol{h}_0$.  We have included an extra condition ${h^\pm_{ab}}^{,b}=0$, providing the Lorenz gauge.



For the fields we define $\mathcal{H}^{aI}_+=\gamma^I_b \mathcal{H}^{ab}_+$  with $\gamma^I_a=-\alpha^{I}_{ab}n^b$ and $n_b=(-1,0,0,0)^T$, with the indexes $I,J, ...$ lowered with the metric $\eta_{IJ}={\rm diag}(1,-1,-1,-1)$. Similarly,  we set $\mathcal{H}^{a\dot I}_-=\gamma^{\dot I}_b \mathcal{H}^{ab}_-$  with $\gamma^{\dot I}_a=-\bar\alpha^{\dot I}_{ab}n^b$, with the indexes $\dot I,\dot J, ...$ lowered with the metric $\eta_{\dot I\dot J}={\rm diag}(1,-1,-1,-1)$. Then, we are able to write the equations for the fields as
\begin{equation}
	{\alpha}_{I}^{a (b}\, \mathcal{H}^{c)I}_{+\ ,a }=0, \qquad \bar{\alpha}_{\dot I}^{a (b}\, \mathcal{H}^{c)\dot I}_{-\ ,a }=0\,,\label{Eq:EOMsdfields}
\end{equation}
that will include Eqs.~\eqref{Eq:Maxwelleqselfdualdiv} and \eqref{Eq:Maxwelleqselfdualrot}, respectively, written as 
\begin{flalign}
	&{\alpha}_{I}^{a (0}\, \mathcal{H}^{j)I}_{+\ ,a }=0\,,\qquad \bar{\alpha}_{\dot I}^{a (0}\, \mathcal{H}^{j)\dot I}_{-\ ,a }=0\,,\\
	&{\alpha}_{I}^{a (i}\, \mathcal{H}^{j)I}_{+\ ,a }=0\,,\qquad \bar{\alpha}_{\dot I}^{a (i}\, \mathcal{H}^{j)\dot I}_{-\ ,a }\,,=0 \,,
\end{flalign}
together with some extra conditions, that become trivial under the gauge choice in \eqref{Eq:TTgauge}. 
Here, $\mathcal{H}^{0I}_+$ and $\mathcal{H}^{0\dot I}_-$ 
act as Lagrange multipliers.  Note that $\gamma^I_a\gamma_I^b=-\delta^b_a=\gamma^{\dot I}_a\gamma_{\dot I}^b\,$, $\,\gamma^I_a\gamma_J^a=-\delta^I_J$ and $\gamma^{\dot I}_a\gamma_{\dot J}^a=-\delta^{\dot I}_{\dot J}$.

These $\alpha$ matrices, with $I$ running from $1$ to $3$, are antisymmetric, invariant under Lorentz transformations, and self-dual $i ^*\alpha^{ab}_I\equiv i\frac{1}{2} \epsilon^{ab}_{\ \ cd}\alpha^{cd}_I=\alpha^{ab}_I$. Furthermore, they satisfy the commutation and anti-commutation relations \cite{Agullo:2018nfv}
\begin{flalign}
	\left\{\alpha_I, \alpha_J\right\} \equiv \alpha^a{ }_{b I} \alpha^{b c}{ }_J+\alpha^a{ }_{b J} \alpha^{b c}{ }_I=2\delta_{I J} \eta^{a c}\,, \\
	\left[\alpha_I, \alpha_J\right] \equiv \alpha^a{ }_{b I} \alpha^{b c}{ }_J-\alpha^a{ }_{b J} \alpha^{b c}{ }_I=2{ }^{+} {\Sigma_{I J}}^{a c}\,,
\end{flalign}
where  ${ }^{+}{\Sigma_{I J}}^{a b}=-i\epsilon_{IJK}\delta^{KL}\alpha_{L}^{ab}$ are the generators of the $(0, 1)$ representation of the Lorentz group and $\delta_{IJ}$ is the Kronecker delta.\footnote{It is  straightforward to derive similar expressions for the conjugate matrices. Here, the anti-commutators will equal ${ }^{-}{\Sigma_{\dot I \dot J}}^{a b}=i\epsilon_{\dot I\dot J\dot K}\delta^{\dot K \dot L}\bar\alpha_{\dot L}^{ab}$, which are the generators of the $(1, 0)$ representation of the Lorentz group.}

These $\alpha_I$ matrices are the spin-1 analogs of the Pauli matrices. Following the reasoning of \cite{Agullo:2018nfv} for electrodynamics, at each spacetime point the fields $\mathcal{H}_\pm$ can be seen as elements of complex vector spaces $V^\pm$ carrying the irreducible representations $(0,1)\otimes(1/2,1/2)$ and $(1,0)\otimes(1/2,1/2)$, respectively, of the Lorentz group. The $\alpha_I$ and $\bar\alpha_{\dot I}$ matrices are isomorphisms to self-dual and anti-self-dual tensors in Minkowski space, and further equip $V^\pm$ with the inner metrics $h_{IJ} = -\delta_{IJ}$ and $h_{\dot I \dot J} = -\delta_{\dot I \dot J}$. These properties have been extensively discussed for electromagnetism in \cite{Agullo:2018nfv} and we refer to this  reference for more technical details. The extension to 4 dimensions by including the additional matrix $\alpha_0^{ab}=\eta^{ab}$ was also discussed in that reference.


The components of the current in Eq.~\eqref{Eq:Currentmaxwellform} become 
\begin{flalign}\label{eq:j0}
	&J^0=\frac{i}{2}\delta\theta\left( \boldsymbol{\mathcal{H}}^-  \cdot \boldsymbol{{h}}^+-\boldsymbol{\mathcal{H}}^+ \cdot \boldsymbol{{h}}^-\right)=\frac{1}{2}\delta\theta(\boldsymbol{h} \cdot \boldsymbol{b}-\boldsymbol{k} \cdot \boldsymbol{e})\,,\\
	&J^i=\frac{i}{2}\delta\theta\left(\boldsymbol{\mathcal{H}}^+ \times \boldsymbol{{h}}^- + \boldsymbol{\mathcal{H}}^-  \times \boldsymbol{{h}}^+\right)^i=\frac{1}{2}\delta\theta(\boldsymbol{e} \times \boldsymbol{h}+\boldsymbol{b} \times \boldsymbol{k})^i\, ,\label{eq:ji}
\end{flalign}
or, equivalently,
\begin{equation}
	J^a=\frac{i}{2}\delta\theta\left(\mathcal{H}_+^{I b}\,\alpha^{a c}_I\, h^-_{bc}-\mathcal{H}_-^{I b}\,\bar{\alpha}^{a c}_I\, h^+_{bc}\right).
\end{equation}


The next step is to write the Lagrangian that describes the dynamics in terms of the self- and anti-self-dual variables. Since the description involving the $\alpha$ matrices resembles that of a massless Dirac equation, we seek a formulation that is also linear in time derivatives. This will render free linearized gravity analogous to Dirac’s theory.


The Lagrangian we consider is
\begin{equation}
	\mathcal{L}_{SD}=-\frac{1}{4}\left(\mathcal{H}_+^{I c}\,\alpha^{a b}_I\, h^-_{bc,a}+ \mathcal{H}_-^{I c}\,\bar{\alpha}^{ab }_I\, h^+_{bc,a}\right)\,,\label{Eq:Lagrangianselfdual}
\end{equation}
where ${h}^\pm_{bc}$ are independent variables, and the quantities $\mathcal{H}^{Ic}_\pm$ are understood as 
\begin{equation}\label{eq:Hic2hab}
{\mathcal{H}^{Ic}_\pm=\pm\, i\eta^{ca}\, \epsilon^{I b d}\,h_{a d, b}^\pm},
\end{equation}
according to Eq.~\eqref{Eq:potentialsdef}, now expressed in the new index notation (see Appendix~\ref{Appendix_B} for a proof that this relation guarantees the equivalence between the field and potential descriptions).

Applying the Euler–Lagrange equations to this Lagrangian yields the desired equations of motion as given in Eq.~\eqref{Eq:EOMsdfields}, where the fields are expressed in terms of Eq.~\eqref{eq:Hic2hab}.\footnote{See also~\cite{Agullo:2018nfv} for the electromagnetic analog.} One can also verify that the Lagrangian is invariant under the duality transformation ${h^\pm_{ab} \rightarrow e^{\mp i \theta} {h}^\pm_{ab}}$, and the corresponding Noether current is given by
\begin{equation}
	J^a=\frac{\partial \mathcal{L}_{SD}}{\partial(h^+_{bc,a})} \delta h^+_{bc}+ \frac{\partial \mathcal{L}_{SD}}{\partial( h^-_{bc,a})} \delta h^-_{bc}=
	\frac{i}{2}\delta\theta\left(\mathcal{H}_+^{I c}\,\alpha^{b a}_I\, h^-_{bc}-\mathcal{H}_-^{I c}\,\bar{\alpha}^{b a}_I\, h^+_{bc}\right)\,,
\end{equation}
that matches the expression obtained in Eqs. \eqref{eq:j0} and \eqref{eq:ji}.

By performing an integration by parts, the expression in Eq.~\eqref{Eq:Lagrangianselfdual} can be rewritten in a symmetric form:
\begin{equation}
	\mathcal{L}=-\frac{1}{8}\left(\mathcal{H}_+^{I c}\,\alpha^{ab}_I\, h^-_{bc,a}-{h}^-_{bc}\,\alpha^{ab}_I\,\mathcal{H}_{+\,,a}^{I c}+ \mathcal{H}_-^{I c}\,\bar{\alpha}^{ab }_I{h}^+_{bc,a}-h^+_{bc}\bar{\alpha}^{ ab}_I\,\mathcal{H}_{-\,,a}^{I c}\, \right)\,.\label{Eq:Lagrangianselfdual1st}
\end{equation}

This action admits a description in terms of spinors and ``gamma''-like matrices. Following~\cite{Agullo:2018nfv} (see Appendix~\ref{Appendix_C} for further details), we introduce
\begin{equation}
	\Psi=\left(\begin{array}{l}
		h^{+}_{bc} \\
		\mathcal{H}_{+}^{Ic} \\
		h_-^{bc} \\
		\mathcal{H}^{-}_{\dot Ic}
	\end{array}\right), \quad \bar{\Psi}=\left(h_{+}^{bc},\mathcal{H}^{+}_{Ic},h^{-}_{bc},
	\mathcal{H}_{-}^{\dot I c}\right), \quad \beta^a=i\left(\begin{array}{cccc}
		0 & 0 & 0 & {\bar\alpha^{\dot I a}}_{\;\;\;\;b} \\
		0 & 0 & - \alpha^{I a}_{\;\;\;\; b}  & 0 \\
		0 & \alpha^{ab}_{ I} & 0 & 0 \\
		-\bar \alpha^{ab}_{\dot I} & 0 & 0 & 0
	\end{array}\right).
\end{equation}
Thus, the action for linearized gravity in a flat background takes the form
\begin{equation}
	S_D=-\frac{1}{4} \int d^4 x  \bar{\Psi} i \beta^a \partial_a \Psi\,,
\end{equation}
which resembles the Dirac theory for a neutral (vanishing electric charge) Majorana 4-spinor. In this case, the two lower components are the complex conjugates of the upper ones.


As discussed in~\cite{Agullo:2018nfv}, one can verify from the algebra of the $\alpha$ matrices that the $\beta^a$ satisfy the Clifford algebra $\text{Cliff}(3,1)$ (see Appendix~\ref{Appendix_C} for details),
\begin{equation}
	\left\{\beta^a, \beta^b\right\}=2\, \eta^{a b} ,
\end{equation}
and $\partial_a \beta^b=0$. These matrices provide a spin-1 representation of the usual Dirac gamma matrices. Furthermore, we can similarly define a chiral matrix as
\begin{equation}
	\beta_5 \equiv \frac{i}{4 !} \epsilon_{abcd} \beta^a \beta^b \beta^c \beta^d=\left(\begin{array}{cccc}
		-\mathbb{I} & 0 & 0 & 0 \\
		0 & -\mathbb{I} & 0 & 0 \\
		0 & 0 & \mathbb{I} & 0 \\
		0 & 0 & 0 & \mathbb{I}
	\end{array}\right)\,,
\end{equation}
which satisfies the standard properties: 
\begin{equation}
	\left\{\beta^\mu, \beta_5\right\}=0, \quad \beta_5^2=\mathbb{I} .
\end{equation}

As with Majorana spinors, the basic dynamical variables in the action are the potentials $\boldsymbol{h}^\pm$. However, from a practical perspective, $\Psi$ and $\bar{\Psi}$ can be treated as independent fields. For instance, varying the action with respect to $\bar{\Psi}$ yields the equations of motion
\begin{equation}~
	i \beta^a \partial_a \Psi=0 \,,
\end{equation}
which resemble a massless Dirac equation. These encompass four coupled equations—one for each component of $\Psi$—corresponding to Eqs.~\eqref{Eq:EOMsdpotentials} and~\eqref{Eq:EOMsdfields}.

Finally, duality rotations of gravitational perturbations can be interpreted as chiral symmetries by noting that
\begin{equation}
	\Psi \rightarrow e^{i \theta \beta_5} \Psi, \quad \bar{\Psi} \rightarrow \bar{\Psi} e^{i \theta \beta_5},
\end{equation}
with $\beta_5$ being the chiral matrix implementing the duality rotation. Again, the Lagrangian remains invariant under this transformation, and the associated conserved current becomes
\begin{equation}
	J^a=\frac{1}{4} \delta\theta  \bar{\Psi} \beta^a \beta_5\Psi=\frac{i}{2}\delta\theta\left(\mathcal{H}_+^{I c}\,\alpha^{b a}_I\, h^-_{b c}-\mathcal{H}_-^{I c}\,\bar{\alpha}^{ba}_I\, h^+_{bc}\right)\, ,\label{Eq:currentDiracflat}
\end{equation}
in agreement with previous expressions.



\section{Extension to curved backgrounds}\label{sec:curvedbd}

We are now interested in extending the previous results to a generic curved background metric. A natural strategy is to follow the standard procedure used for Dirac spin-$1/2$ fields, which has already been successfully applied in \cite{Agullo:2018nfv} to spin-1 fields. 

First of all, we promote the Minkowski metric $\eta_{ab}$ to the general curved metric $g_{\mu\nu}(x)$. The second step consists in introducing an orthonormal tetrad field, or vierbein,  $e^\mu_a (x)$, which relates the curved and flat metrics via $g_{\mu\nu}(x)=\eta_{ab}e^a_{\mu}(x)e^b_{\nu}(x)$. The curved spacetime $\alpha$-matrices are then obtained from their flat spacetime counterparts according to
\begin{equation}
	{\alpha_I^{\mu \nu}(x)=e_a^\mu(x) e_b^\nu(x) }\alpha_I^{a b}\,.
\end{equation}
The curved spacetime properties of the $\alpha$-matrices follow closely the discussion in \cite{Agullo:2018nfv}. Now, to define the covariant derivative $\nabla_\mu$ we adopt standard arguments (see, e.g., \cite{Ashtekar:1991hf}). This is, we demand compatibility with the isomorphism, namely that $\nabla_\beta\alpha_{I}^{\mu\nu}=0$. This ensures consistency with the local Lorentz structure. 
%
%
The action of the covariant derivative on the fields $\mathcal{H}^{I\nu}_\pm$ can then be expressed in terms of a 1-form spin connection $w_\mu^{ab}$, as 
\begin{flalign}
 \nabla_\mu \mathcal{H}_+^{I\nu}= \partial_\mu \mathcal{H}_+^{I\nu} - \frac{1}{2} (w_\mu)_{ab} \left[ ^+\Sigma^{ab}\right]^I_{\ J} \mathcal{H}^{J\nu}_+ +\Gamma^\nu_{\ \mu\rho}\mathcal{H}^{I\rho}_+\, ,
 \\
 \nabla_\mu \mathcal{H}_-^{I\nu}= \partial_\mu \mathcal{H}_-^{I\nu} - \frac{1}{2} (w_\mu)_{ab} \left[ ^-\Sigma^{ab}\right]^I_{\ J} \mathcal{H}^{J\nu}_- +\Gamma^\nu_{\ \mu\rho}\mathcal{H}^{I\rho}_-\, ,
\end{flalign}
where $ ^\pm\Sigma$ are the generators of the $(0,1)$  and $(1,0)$ representation of the Lorentz Lie algebra, respectively. The 1-form connection is obtained from the vierbein as 
\begin{equation}
 	(w_\mu)^a_{\ b}=e^a_\alpha\partial_\mu e^\alpha_b+e^a_\alpha e^\beta_b\Gamma^\alpha_{\ \mu\beta}\, ,
\end{equation}
where $\Gamma^\alpha_{\ \mu\beta}$ are the Christoffel symbols of the Levi-Civita connection. This formalism provides a consistent extension of the linearized description of gravity to a generic background metric. 

The equations of motion for the potentials--i.e., the metric perturbations--take the covariant form 
\begin{equation}
	\bar{\alpha}_{(I}^{\mu \nu}\, h_{\lambda)\nu;\mu}^{+}=0\,,
	\qquad {\alpha}_{(I}^{\mu \nu}\, h_{\lambda)\nu;\mu}^{-}=0\,,\label{Eq:EOMsdpotentialscurved}
\end{equation}
and the dynamics with the fields is given by 
\begin{equation}
	{\alpha}_{I}^{\mu (\nu}\, \mathcal{H}^{\lambda)I}_{+\ ;\mu }=0, \qquad \bar{\alpha}_{I}^{\mu (\nu}\, \mathcal{H}^{\lambda)I}_{-\ ;\mu }=0\,,\label{Eq:EOMsdfieldscurved}
\end{equation}
where the semicolon denotes the covariant derivative. The relation between the (anti) self-dual Fierz tensors and the potentials, as introduced earlier in Eq.\eqref{Eq:fierz_fields}, leads directly to the covariant field equations 
$F^{\mu\nu\lambda}_\pm\,_{;\mu}=0$. Additionally, the relation between fields and potentials generalizes naturally to:
\begin{equation}
	\mathcal{H}^{ I\lambda}_\pm=i\, \epsilon^{I\mu\nu}h_{\,\nu}^{\pm\, \lambda}\,_{;\mu}\, ,\label{Eq:relationfieldspotentials}
\end{equation}
as detailed in Appendix \ref{Appendix_B}.

All of this culminates in a generalized Dirac-like action for gravitational perturbations propagating on a generic curved background:
\begin{equation}
	S_D=-\frac{1}{4} \int d^4 x \sqrt{-g} \;  \bar{\Psi} i \beta^\mu \nabla_\mu \Psi\,,
\end{equation} 
with the field content and matrices given by:
\begin{equation}\label{eq:psibetamu}
	\Psi=\left(\begin{array}{l}
		h^{+}_{\nu \lambda} \\
		\mathcal{H}_{+}^{I\lambda} \\
		h_{-}^{\nu \lambda} \\
		\mathcal{H}^{-}_{\dot I\lambda}
	\end{array}\right), \quad \bar{\Psi}=\left(h^{+}_{\nu \lambda},\mathcal{H}_{+}^{I\lambda},h^{-}_{\nu \lambda},
	\mathcal{H}_{-}^{I\lambda}\right)\,, \quad     \beta^\mu = i\left(\begin{array}{cccc}
		0 & 0 & 0 & {\bar\alpha^{\dot I \mu}}_{\;\;\;\;\nu} \\
		0 & 0 & - \alpha^{I \mu}_{\;\;\;\; \nu} & 0 \\
		0 & \alpha^{\mu\nu}_{ I} & 0 & 0 \\
		-\bar \alpha^{\mu\nu}_{\dot I} & 0 & 0 & 0
	\end{array}\right).
\end{equation}

  The  $\beta_\mu$ matrices satisfy the Clifford algebra  in  curved spacetime, $	\left\{\beta^\mu, \beta^\nu\right\}=2\, g^{\mu\nu},$ as well as $\nabla_\nu\beta^\mu=0$. Hence, the dynamics for the {metric perturbations} is governed by the covariant, massless Dirac-like equation:
\begin{equation}~
	i \beta^\mu \nabla_\mu \Psi=0 \,.
\end{equation}
On the other hand, the conserved current associated with the continuous duality symmetry generalizes Eq. \eqref{Eq:currentDiracflat} to:
\begin{equation}
	J^\mu=\frac{1}{4} \delta\theta\    \bar{\Psi} \beta^\mu \beta_5\Psi=-\frac{i}{2}\delta\theta\left(\mathcal{H}_+^{I \lambda}\,\alpha^{\mu\nu}_I\, h^-_{\mu\lambda}-\mathcal{H}_-^{I \lambda}\,\bar{\alpha}^{\mu\nu}_I\, h^+_{\mu\lambda}\right)\,.
\end{equation}

Using the commutation properties satisfied by the $\beta^\mu$ matrices, a second-order, Klein-Gordon-like equation holds for $\Psi$:
  \begin{equation}
  	\left(-i \beta^\mu \nabla_\mu\right) i \beta^\nu \nabla_\nu \Psi=\left(\beta^{(\mu} \beta^{\nu)}+\beta^{[\mu} \beta^{\nu]}\right) \nabla_\mu \nabla_\nu \Psi=(\square+\mathcal{Q}) \Psi=0,
  \end{equation}
  where
  \begin{equation}
  	\mathcal{Q} \Psi \equiv \frac{1}{2} \beta^{[\mu} \beta^{\nu]} W_{\mu \nu} \Psi,\label{Eq:defQ}
  \end{equation}
with
\begin{equation}\label{Eq:defWmunu}
W_{\mu \nu} \Psi \equiv\left[\nabla_\mu, \nabla_\nu\right] \Psi=R_{\mu \nu \sigma \delta}\left(\begin{array}{l}
		(\Sigma^{\sigma \delta}){}_{\alpha \beta}{}^{\gamma\epsilon}h^{+}_{\gamma \epsilon} \\
		^{+}(\Sigma^{\sigma \delta}){}^{I\alpha}{}_J{}_{\gamma}\mathcal{H}_{+}^{J\gamma} \\
		(\Sigma^{\sigma \delta}){}^{\alpha \beta}{}_{\gamma\epsilon}h_{-}^{\gamma \epsilon} \\
		^{-}(\Sigma^{\sigma \delta}){}_{\dot I\alpha}{}{}^{\dot J\gamma}\mathcal{H}^{-}_{\dot J\gamma}
	\end{array}\right),
\end{equation}
or, in matrix form,
\begin{equation}
	W_{\mu \nu} \Psi = R_{\mu \nu \sigma \delta}\left(\begin{array}{cccc}
		\Sigma^{\sigma \delta} & 0 & 0 & 0 \\
		0 & {}^+\Sigma^{\sigma \delta} & 0 & 0 \\
		0 & 0 & \Sigma^{\sigma \delta} & 0 \\
		0 & 0 & 0 & {}^-\Sigma^{\sigma \delta}
	\end{array}\right) \Psi .\label{Eq:defWmunu-mat}
\end{equation}
Similarly to   electrodynamics \cite{Agullo:2018nfv}, 
here $\Sigma^{\sigma \delta}$ is  shorthand for
\begin{equation}
(\Sigma^{\sigma \delta}){}_{\alpha \beta}{}^{\gamma\epsilon}=4\delta^{[\sigma|}_{(\alpha}
\delta^{(\epsilon}_{\beta)}g^{\gamma)|\delta]}   \, ,
\end{equation}
(symmetric in $(\alpha,\beta)$ and $(\gamma,\epsilon)$ and antisymmetric in $(\sigma,\delta)$), which is the generator of the $(1,1)$ (real) representations of the Lorentz group. Moreover, $ ^+\Sigma^{\sigma \delta}$,  $ ^-\Sigma^{\sigma \delta}$ correspond to
\begin{equation}\label{eq:sigmapHil}
^{+}(\Sigma^{\sigma \delta}){}^{I\alpha}{}_J{}_{\gamma}=
-\frac{i}{2} \delta^{\alpha}_{\gamma} \epsilon^I{}_{JK} \alpha^{K\sigma\delta},\quad 
^{-}(\Sigma^{\sigma \delta}){}^{\dot I\alpha}{}_{\dot J}{}_{\gamma}= 
\frac{i}{2} \delta^{\alpha}_{\gamma} \epsilon^{\dot I}{}_{\dot J\dot K} \bar\alpha^{\dot K\sigma\delta},
\end{equation}
which are generators of the $(0,1)\bigotimes(1/2,1/2)$ and
$(1,0)\bigotimes(1/2,1/2)$ representations, respectively, of the Lorentz group. More details can be found in Appendix \ref{Appendix_D}.
 
Putting everything together, the operator $\cal Q$ in Eq. \eqref{Eq:defQ}, with the definition of $\beta^{[\mu} \beta^{\nu]}$ given in Eq. \eqref{eq:betamunu}, produces 
  \begin{equation}
  	\mathcal{Q} \Psi \equiv \frac{1}{2} R_{\mu \nu \sigma \delta}\left(\begin{array}{l}
		4 \, ^-P^{\mu\nu\;\lambda}_{\;\;\;\;\alpha}(\Sigma^{\sigma \delta}){}_{\lambda\beta}{}^{\gamma\epsilon}h^{+}_{\gamma \epsilon} \\
		(-1) \, ^+M_{\;\;J}^{I\;\; \mu\nu}{}^{+}(\Sigma^{\sigma \delta}){}^{J\alpha}{}_K{}_{\gamma}\mathcal{H}_{+}^{K\gamma} \\
		4 \, ^+P^{\mu\nu\alpha}_{\;\;\;\;\;\;\;\lambda}(\Sigma^{\sigma \delta}){}^{\lambda\beta}{}_{\gamma\epsilon}h_{-}^{\gamma \epsilon} \\
		(-1) \, ^-M_{\dot I}^{\;\;\dot J \mu\nu}{}^{-}(\Sigma^{\sigma \delta}){}_{\dot J\alpha}{}{}^{\dot K\gamma}\mathcal{H}^{-}_{\dot K\gamma}
	\end{array}\right).\label{Eq:defQ}
  \end{equation}

  As a final remark, note that the spinor structure is preserved under these curvature corrections.
The resulting spinors (in components) are of the form
  \begin{equation}
  	\mathcal{Q} \Psi \equiv \left(\begin{array}{l}
		{\tilde\Psi}^{1}_{\alpha \beta} \\
		{\tilde\Psi}_{2}^{\alpha I} \\
		\tilde{\Psi}_{3}^{\alpha \beta} \\
		{\tilde\Psi}^{4}_{\alpha \dot I}
	\end{array}\right)\, ,
  \end{equation}
 and remain in the same representation space.
Specifically, in the definitions above of the $\beta^\mu$ matrices, some index contractions have been left implicit for compactness; the explicit index structure is recovered through the action of the operators defined above.
  

\section{Quantum anomaly}
\label{sec:anomaly}

We finally compute the vacuum expectation value of the divergence of the classically conserved current, $\langle\nabla_\mu J_D^\mu\rangle$. A non-vanishing result would imply that the vacuum expectation value of the charge $Q_D$ is not conserved in time. Since we have successfully described gravitational waves using variables analogous to those in electromagnetism, and reformulated them in a spinorial formalism, we can now rely on previous results in the literature \cite{Agullo:2018nfv}. We reproduce the anomaly calculation through a direct computation, in which ultraviolet divergences are identified and subtracted in a covariant and self-consistent manner. 
The only significant difference in our gravitational case arises from the presence of additional vector indices in the spinor components.

The quantity of interest is quadratic in the field variables and therefore exhibits ultraviolet (UV) divergences, as usual in quantum field theory. To obtain finite results, we must renormalize the expectation value by subtracting the fourth-order DeWitt-Schwinger adiabatic expansion \cite{Parker:2009uva}:
%
\begin{equation}
	\left\langle\nabla_\mu J^\mu\right\rangle_{\mathrm{ren}}=\left\langle\nabla_\mu J^\mu\right\rangle-\left\langle\nabla_\mu J^\mu\right\rangle_{\mathrm{Ad}(4)}\, .
\end{equation}
The renormalization proceeds by expressing $\left\langle\nabla_\mu J^\mu\right\rangle$ in terms of the Feynman two-point function $S(x,x') = -i\langle T \Psi(x) \bar{\Psi}(x') \rangle$, and subtracting its adiabatic expansion $S(x,x')_{\text{Ad}(4)}$. Finally, one takes the coincidence limit $x \rightarrow x'$. To regularize spurious infrared divergences, we modify the wave equation as $(i\beta^\mu \nabla_\mu + m)\Psi = 0$, introducing an auxiliary mass parameter $m > 0$, to be sent to zero at the end of the calculation.

Using the $\beta^\mu$ commutation relations, we find:
\begin{equation}
	\begin{aligned}
		\nabla_\mu J^\mu(x) & =\nabla_\mu\left[\frac{1}{4}  \bar{\Psi}(x) \beta^\mu \beta_5 \Psi(x)\right]=-\frac{i}{4}  \left(\bar{\Psi}(x) \overleftarrow{D} \beta_5 \Psi(x)-\bar{\Psi}(x) \beta_5 \vec{D} \Psi(x)\right)\\
		& =\lim _{\substack{m \rightarrow 0 \\
				x \rightarrow x^{\prime}}} \frac{-i\,m}{2}  \bar{\Psi}(x) \beta_5 \Psi\left(x^{\prime}\right)=\lim _{\substack{m \rightarrow 0 \\
				x \rightarrow x^{\prime}}} \frac{-i\,m}{2} \operatorname{Tr}\left[\beta_5 \Psi(x) \bar{\Psi}\left(x^{\prime}\right)\right]\,,
	\end{aligned}
\end{equation}
where $D = i \beta^\mu \nabla_\mu$. Taking the expectation value in an arbitrary vacuum state gives:
\begin{equation}
	\left\langle\nabla_\mu J^\mu\right\rangle=\lim _{\substack{m \rightarrow 0 \\ x \rightarrow x^{\prime}}} \frac{1}{2} m \operatorname{Tr}\left[\beta_5 S\left(x, x^{\prime}, m\right)\right].
\end{equation}
Then, the renormalized expectation vacuum becomes
\begin{equation}
	\left\langle\nabla_\mu J^\mu\right\rangle_{\text {ren }}=\lim _{\substack{m \rightarrow 0 \\ x \rightarrow x^{\prime}}} \frac{m}{2} \operatorname{Tr}\left[\beta_5\left(S\left(x, x^{\prime}, m\right)-S\left(x, x^{\prime}, m\right)_{\operatorname{Ad}(4)}\right)\right].
\end{equation}
Here, $S(x,x',m)$ encodes the information about the vacuum state, while $S(x,x',m)_{\mathrm{Ad}(4)}$ accounts for the universal asymptotic structure of the Wightman bi-distribution as the two points get close, ensuring the subtraction of state-independent UV divergences. Following \cite{Parker:2009uva}, we write $S\left(x, x^{\prime}, m\right)_{\mathrm{Ad}(4)}=\left[(D-m) G\left(x, x^{\prime}, m\right)\right]_{\mathrm{Ad}(4)}$, where
\begin{equation}
	G\left(x, x^{\prime}, s\right) \sim \frac{\hbar \Delta^{1 / 2}\left(x, x^{\prime}\right)}{16 \pi^2} \sum_{k=0}^{\infty} E_k\left(x, x^{\prime}\right) \int_0^{\infty} d \tau e^{-i\left(\tau m^2+\frac{\sigma\left(x, x^{\prime}\right)}{2 \tau}\right)}(i \tau)^{(k-2)}.
\end{equation}
Here, $\sigma(x,x')$ is half the squared geodesic distance between $x$ and $x'$, $\Delta^{1/2}(x,x')$ is the Van Vleck–Morette determinant, and $E_k(x,x')$ are the DeWitt coefficients, built from the background geometry. Namely, each $E_k(x,x')$ is obtained from the metric and its first $2k$ derivatives.


We are interested in the coincidence limit $x \to x'$. Because of the underlying symmetry in the classical theory, the bare contribution vanishes for all vacuum states, as it only depends on the field modes, which verify the field equations exactly. Hence, the entire anomaly arises from the subtraction term:
As a result, $\left\langle\nabla_\mu J^\mu\right\rangle_{\text {ren }}$ arises simply from $S(x, x' , m)_{Ad(4)}$. The expressions for $k = 0,1,2$ are given by \cite{Parker:2009uva} (see Eqs. (5.57)-(6.60))
\begin{equation}
    \begin{aligned}
		E_0(x) & = \mathbb{I}, \\ 
        E_1(x) & = \frac{1}{6}R\mathbb{I}-\mathcal{Q},\\
        E_2(x) & =\left[-\frac{1}{30} \square R+\frac{1}{72} R^2-\frac{1}{180} R_{\mu \nu} R^{\mu \nu}+\frac{1}{180} R_{\alpha \beta \mu \nu} R^{\alpha \beta \mu \nu}\right] \mathbb{I} \\
		& +\frac{1}{12} W_{\mu \nu} W^{\mu \nu}+\frac{1}{2} \mathcal{Q}^2-\frac{1}{6} R \mathcal{Q}+\frac{1}{6} \square \mathcal{Q},
    \end{aligned}
\end{equation}
with $W_{\mu\nu}$ and $\mathcal{Q}$ given in Eqs. \eqref{Eq:defWmunu} and \eqref{Eq:defQ}, respectively. All other terms can be easily deduced (see Appendix \ref{Appendix_D}). 
We will see that only the terms with $k = 2$  produce a non-vanishing result. Additionally, terms involving derivatives of $E_2 (x, x' )$ must be disregarded because they involve five derivatives of the metric and hence are of the fifth adiabatic order. One can easily see that $\operatorname{Tr}\left[\beta_5 E_0(x, x)\right]=\operatorname{Tr}\left[\beta_5\right]=0$. Besides, 
\begin{equation}
\operatorname{Tr}\left[\beta_5 E_1(x, x)\right] = \frac{1}{6}R\operatorname{Tr}\left[\beta_5\right]-\operatorname{Tr}\left[\beta_5 {\cal Q}\right].
\end{equation} 
The first addend is zero. The second one is $\operatorname{Tr}\left[\beta_5 {\cal Q}\right]=-\frac{i}{4}\epsilon_{\mu\nu\rho\sigma}R^{\mu\nu\rho\sigma}=0$ because of the first Bianchi identity of the Riemann tensor. Regarding $\operatorname{Tr}\left[\beta_5 E_2(x, x)\right]$, and following similar arguments, the only non-vanishing terms are
\begin{eqnarray}
    &&\operatorname{Tr}\left[\beta_5 W_{\mu\nu}W^{\mu\nu}\right]= i\epsilon_{\mu\nu\rho\sigma}  R^{\alpha\beta\mu\nu} R_{\alpha\beta}{}^{\rho\sigma}\, ,    \\
    &&\operatorname{Tr}\left[\beta_5 {\cal Q}^2\right] = -i\frac{5}{8}\epsilon_{\mu\nu\rho\sigma}  R^{\alpha\beta\mu\nu} R_{\alpha\beta}{}^{\rho\sigma}.
\end{eqnarray}
Some of the details of this calculation can be found in Appendix \ref{Appendix_D}. One finds 
\begin{equation}
	\left.\operatorname{Tr}\left[\beta_5 E_2(x, x)\right)\right]=\operatorname{Tr}\left[\frac{1}{12}\beta_5 W_{\mu\nu}W^{\mu\nu} +\frac{1}{2}\beta_5 {\cal Q}^2\right]=-i \frac{11}{24} R_{\alpha \beta \mu \nu}{ }^{\star} R^{\alpha \beta \mu \nu}\, ,
\end{equation}
where ${ }^{\star} R^{\alpha \beta \mu \nu}=\frac{1}{2}\epsilon^{\alpha\beta\sigma\rho} R_{\sigma\rho}{}^{\mu \nu}$ is the dual of the Riemann tensor. Finally, we obtain 
\begin{equation}
	\left\langle\nabla_\mu J^\mu\right\rangle_{\text {ren }}=-i\frac{\hbar}{32\pi^2}{\rm Tr}\left[\beta_5E_2(x, x)\right]=-\hbar\frac{11}{768\pi^2}  R_{\alpha \beta \mu \nu}{ }^{\star} R^{\alpha \beta \mu \nu}.
\end{equation}	

This is the main result of our paper. 

It is useful to compare this anomaly with the curvature term omitted in the equations of motion (see Appendix~\ref{Appendix_E}). Variations of the Fierz-Pauli action \eqref{Eq:fierz_fields} yield the field equation:
\begin{equation}
	\nabla^\lambda \nabla_\lambda h_{\mu \nu}- R_{\nu \lambda \sigma \mu} h^{\lambda \sigma}=0,
\end{equation}
while the linearized Einstein-Hilbert action in Eq. \eqref{eq:Eq:lin-HE} gives:
\begin{equation}
	\nabla^\lambda \nabla_\lambda h_{\mu \nu}- 2R_{\nu \lambda \sigma \mu} h^{\lambda \sigma}=0.
\end{equation}
As we can see, the two descriptions are not exactly equivalent in curved spacetimes. However, in the geometric optics approximation,  $\lambdabar \ll L_B$ \cite{Maggiore:2007ulw} ---where $\lambdabar$ is the characteristic gravitational-perturbation wavelength and  $L_B$  the characteristic length scale of the background curvature--- the neglected curvature term scales as ${R}_{\mu \lambda  \sigma\nu} {h}^{\lambda \sigma}\sim O(|h_{\mu\nu}|/L_B^2)$, while the dominant kinetic term scales as $\nabla^\lambda \nabla_\lambda {h}_{\mu \nu}\sim O(|h_{\mu\nu}|/\lambdabar^2)$, justifying the approximation.


Now, the anomaly scales as $\kappa \hbar R_{\alpha\beta\mu\nu}{}^{\star}R^{\alpha\beta\mu\nu} \sim \ell_{Pl}^2 / L_B^4$ (with $\kappa=8\pi G$). If we require the curvature term in the field equations to be much smaller than the anomaly  in the current,  $|h_{\mu\nu}| / L_B^2\ll \ell_{Pl}^2 / L_B^4$, or equivalently $|h_{\mu\nu}| \ll \ell_{Pl}^2 / L_B^2$. On the other hand, the semiclassical approximation requires $|h_{\mu\nu}| \simeq \ell_{Pl}^2 \lambdabar^2 / L_B^4\ll 1$ for $\ell_{Pl} / L_B\ll 1$ (which is in turn required for gravity to remain classical). When both are combined, we get $\lambdabar \ll L_B$ as consistency condition. 
%
%
Therefore, the approximations adopted here are justified provided  $\lambdabar/L_B\ll 1$, $\ell_{Pl} / L_B\ll 1$ and $h \simeq \ell_{Pl}^2 \lambdabar^2 / L_B^4$.

We also want to note that an alternative formulation could have been achieved with the variables $h^+_{\mu J}=h^+_{\mu\lambda}\gamma^{\lambda}_J$ and $\mathcal{H}^{IJ}_+=\mathcal{H}^{I\lambda}_+\gamma_{\lambda}^J$, and similarly  $h^-_{\mu \dot J}=h^-_{\mu\lambda}\gamma^{\lambda}_{\dot J}$ and $\mathcal{H}^{\dot I\dot J}_-=\mathcal{H}^{\dot I\lambda}_-\gamma_{\lambda}^{\dot J}$. In a flat background, since $\partial_a\gamma^b_I=\alpha_I^{bc}\partial_a n_c=0$ (and similarly for the conjugate $\gamma^b_{\dot I}$), the equations of motion coincide and the action can easily be rewritten with these quantities. 
On the other hand, in a curved background, the equations of motion have similar structure but describe different quantities as $\nabla_\mu \gamma^\nu_I=\alpha_I^{\nu\rho}\nabla_\mu n_\rho\ne 0$ (and also for the conjugate). Nevertheless, $\nabla_\mu n_\rho$ defines the extrinsic curvature which, assuming it has $L_B$ as characteristic scale, can be neglected against the derivatives of the gravitational perturbations with characteristic length $\lambdabar$. Hence, within this approximation, both theories have the same equations of motion and satisfy the same dual symmetry. The conserved currents and charges coincide. The anomaly in the current is also the same. The reason being that the only change we are making is a relabeling of a spacetime index by either a self-dual or anti-self-dual indexes with the matrices $\gamma_{\lambda}^J$ and $\gamma_{\lambda}^{\dot J}$. Hence, both descriptions are trivially related. We keep the first description for convenience. 

\section{Conclusions and final remarks}\label{sec:concl}



This paper has explored the classical and quantum aspects of duality rotations for perturbative gravitational waves propagating in general curved spacetimes, drawing analogies with electric-magnetic duality in Maxwell theory and chiral symmetries in fermionic field theories. Specifically, we considered the vacuum Einstein equations in the linearized regime and identified the electric and magnetic components of the flat metric perturbations. These components exhibit an electric-magnetic duality symmetry analogous to that of Maxwell theory. We identified a conserved Noether current, whose associated charge quantifies the difference in intensity between the right- and left-handed circularly polarized components of the gravitoelectromagnetic field—namely, its self-dual and anti-self-dual parts. Notably, this conservation law holds even when the perturbations evolve on arbitrary classical curved backgrounds.


To delve deeper into this symmetry, we reformulated the theory in terms of self-dual and anti-self-dual variables, inspired by prior developments in electrodynamics. This formulation enabled us to cast the dynamics into a Dirac-like equation, thereby establishing a direct correspondence between the helicity of gravitational waves and the chirality of some spinor fields. Within this framework, duality rotations of the gravitational perturbations are naturally interpreted as chiral transformations, implemented via a generalized “chiral” matrix $\beta_5$.


The central result of our work is that quantum fluctuations break this classical duality symmetry when gravitational waves propagate in curved spacetimes. We computed the vacuum expectation value of the divergence of the classically conserved current. To handle the associated ultraviolet divergences, we employed heat-kernel renormalization using the DeWitt-Schwinger asymptotic expansion. While the symmetry is preserved in flat spacetime, in curved backgrounds we find a non-vanishing anomaly. This quantum effect spoils the conservation of the axial current and thus breaks the classical duality symmetry. The result generalizes the notion of chiral anomalies—originally discovered in fermionic and spin-one fields—to massless spin-two fields. A direct physical implication is the appearance of a net polarization of gravitational wave quanta in curved spacetimes.


We conclude by noting that the Maxwell-like description of linearized gravity developed here differs from the standard derivation of gravitational waves directly from the Einstein-Hilbert action in curved spacetime. This discrepancy originates from the order in which the transverse-traceless (TT) gauge conditions are imposed and total derivatives are added to reorganize the Lagrangian. However, both descriptions coincide in the geometric optics approximation, which applies when the wavelength of the perturbation is much smaller than the typical curvature scale of the background. Under appropriate semiclassical conditions, this approximation validates the use of the Fierz-Pauli description for gravitational perturbations decoupled from the background.

Future directions may include exploring gravitoelectromagnetic formulations of the Weyl tensor, investigating its connection with the Lanczos potential and its self- and anti-self-dual decomposition, and developing a parallel description using Ashtekar’s self-dual variables. These formulations may offer further insight into the role of duality and chiral structures in quantum gravity.


\acknowledgments

The authors would like to thank Ivan Agullo and Mar Bastero-Gil for helpful discussions. Financial support is provided by the Spanish Government through the projects PID2020-119632GB-I00, and PID2019-105943GB-I00 (with FEDER contribution). 
ATM is supported by FCT CERN grant with DOI identifier 10.54499/2024.00252.CERN.
ADR acknowledges support through {\it Atraccion de Talento Cesar Nombela} grant No 2023-T1/TEC-29023, funded by Comunidad de Madrid (Spain), and   financial support  via the Spanish Grant  PID2023-149560NB-C21, funded by MCIU/AEI/10.13039/501100011033/FEDER, UE.

\appendix

\section{Properties of $\alpha$ matrices}\label{Appendix_A}

The $\alpha^{ab}_I$ matrices with $I=1,2,3,$ have the following properties:
\begin{eqnarray}
\alpha_{ab I} \, \alpha^{ab}_{J} &=& 4h_{IJ}=-4\delta_{IJ}\, ,\quad \bar\alpha_{ab \dot I} \, \bar \alpha^{ab}_{\dot J}  = 4h_{\dot I\dot J}= -4\delta_{\dot I\dot J}\, , \label{prop1a}\\
h^{IJ}\alpha^{ab}_{I} \,  \alpha_J^{cd}   &=&  4 \, ^+P^{abcd} \, ,\quad
h^{\dot I\dot J}\bar \alpha^{ab}_{\dot I} \,  \bar \alpha_{\dot J}^{cd}=4 \, ^-P^{acbd} \, ,\label{prop2a}\\
\alpha_{ab I} \, \bar \alpha^{ab}_{\dot J} &=& 0 \label{prop3a}\, , \\
\alpha^{a}_{\ b I} \, \alpha^{cb}_{ J} & = &  \eta_{IJ}{\eta}^{ac}- \, i\epsilon_{IJK}h^{KL}\, {\alpha_L^{ab}}\, , \quad 
\bar \alpha^{a}_{\ b \dot I} \, \bar \alpha^{cb}_{ \dot J}  =   \eta_{\dot I\dot J}{\eta}^{ac}+i\epsilon_{\dot I\dot J\dot K}h^{\dot K \dot L}\, {\bar \alpha^{ab}}_{\dot L} \label{prop4a}\, ,
\end{eqnarray}
where $^\pm P^{abcd}=\frac{1}{4}(\eta_{ac}\eta_{bd}-\eta_{ad}\eta_{bc}\pm i \epsilon_{abcd})$ are the projectors on self-dual and anti-self-dual tensors in Minkowski spacetime, respectively, and $\epsilon_{IJK}$ and $\epsilon_{\dot I\dot J\dot K}$ are totally antisymmetric (Levi-Civitta) symbols in the corresponding self- and anti-self-dual sectors. 

Now, the $\alpha^{ab}_I$ matrices with $I=0,1,2,3\,$, (recalling that $n^{I}\alpha^{ab}_I=\eta^{ab}$) introduced in \eqref{eq:alphas} have the following properties:
\begin{eqnarray}
\alpha_{ab I} \, \alpha^{ab}_{J} &=& 4\eta_{IJ}\, ,\quad \bar\alpha_{ab \dot I} \, \bar \alpha^{ab}_{\dot J}  = 4\eta_{\dot I\dot J}\, , \label{prop1}\\
\eta^{IJ}\alpha^{ab}_{I} \,  \alpha_J^{cd}   &=&  4 \, ^+P^{abcd} +{\eta}^{ab}{\eta}^{cd}\, ,\quad
\eta^{\dot I\dot J}\bar \alpha^{ab}_{\dot I} \,  \bar \alpha_{\dot J}^{cd}=4 \, ^-P^{acbd} +{\eta}^{ac}{\eta}^{bd}\, ,\label{prop2}\\
\alpha_{ab I} \, \bar \alpha^{ab}_{\dot J} &=& 4n_I n_{\dot J} \label{prop3}\, , \\
\alpha^{a}_{\ b I} \, \alpha^{cb}_{ J} & = &  \eta_{IJ}{\eta}^{ac}- \, ^+M_{IJ}^{ac}\, , \quad 
\bar \alpha^{a}_{\ b \dot I} \, \bar \alpha^{cb}_{ \dot J}  =   \eta_{\dot I\dot J}{\eta}^{ac}- \, ^-M_{\dot I\dot J}^{ac} \label{prop4}\, ,
\end{eqnarray}
where $\eta_{IJ}={\rm diag}(1,-1,-1,-1)=\eta_{\dot I\dot J}$ are the metrics for the internal indexes $I,J,\ldots $ for self-dual variables and $\dot I,\dot J,\ldots $ for anti-self-dual variables, respectively; they fulfill $\eta_{IJ}=h_{IJ}+n_{I}n_J$, and similarly $\eta_{\dot I\dot J}=h_{\dot I\dot J}+n_{\dot I}n_{\dot J}$; $^+M_{IJ}^{ab}=\,  \, i\epsilon_{IJK}h^{KL}\, {\alpha_L^{ab}} +2\,  \alpha^{ab}_{K}h^K_{[I} n_{J]} $ and $^-M_{\dot I \dot J}^{ab}=-i\epsilon_{\dot I\dot J\dot K}h^{\dot K \dot L}\, {\bar \alpha^{ab}}_{\dot L} +2\,  \bar\alpha^{ab}_{\dot K}h^K_{[\dot I} n_{\dot J]} $. Actually, all the above expressions can be easily derived out of Eqs. \eqref{prop1a}-\eqref{prop4a}.

It is also convenient to introduce the maps 
\begin{equation}\nonumber
 \gamma^a_I=\alpha_{I}^{ab}n_b, \quad \gamma^a_{\dot I}=\bar\alpha_{\dot I}^{ab}n_b, \quad \gamma_I^{\dot J}=\gamma_I^a\gamma_a^{\dot J}. \label{eq:gammas}
\end{equation}
The first and second ones map self- and anti-self-- dual fields into spacetime vectors, respectively, while the latter maps self-dual into anti-self--dual fields. 

In this way, we can also introduce the totally antisymmetric, ``purely spatial'' tensor with mixed indices 
\begin{equation}
{\epsilon_I}^{ab}:= \frac{1}{2i} (\alpha^{ab}_I-\bar \alpha_{\dot J}^{ab}\, \gamma^{\dot J}_I) = \gamma_I^{c}\, n^{d}\, {\epsilon_{dc}}^{ab}.\label{Eq:epsilonIab}
\end{equation}
Here, capital Latin indexes run from 1 to 3. 

\section{Equivalence in description between $\mathcal{H}$ and $h$}\label{Appendix_B}
We now discuss the equivalence between the equations of motion in terms of the fields $\mathcal{H}^{I\lambda}_\pm$ and the potentials $h^\pm_{\nu\lambda}$. We detail the equivalence with self-dual quantities but the derivation for anti-self-dual variables is analogous and obtained by complex conjugation.

Let us start from the potentials e.o.m., $\bar{\alpha}^{\mu\nu}_{\dot{I}} \nabla_\mu h^{+\,\lambda}_{\,\nu}=0$. 
Considering the identity in Eq. \eqref{Eq:epsilonIab} the previous equation implies
\begin{equation}
     2 i\  \epsilon^{I\mu\nu} \nabla_\mu h^{+\,\lambda}_{\,\nu}= \alpha^{I \mu\nu} \nabla_\mu h^{+\,\lambda}_{\,\nu}.
\end{equation}
From the relation between fields and potentials, see \eqref{Eq:relationfieldspotentials}, we have 
$    \mathcal{H}_+^{I\lambda}=\frac{1}{2}\alpha^{I\mu\nu} \nabla_\mu h^{+\,\lambda}_{\,\nu}.$ Then applying $\alpha_I^{\delta\rho}\nabla_\rho$ we find 
\begin{equation}
    \alpha_I^{\delta\rho}\nabla_\rho\mathcal{H}_+^{I\lambda}=\frac{1}{2}\alpha_I^{\delta\rho}\nabla_\rho\alpha^{I\mu\nu} \nabla_\mu h^{+\,\lambda}_{\,\nu}=-2 \nabla_\rho\nabla^{[\delta} h_+^{\rho]\lambda},
\end{equation}
where we have used Eq.\eqref{prop2} in the simplification. Finally, since $h_+^{\rho\lambda} $ satisfy the second order equations we have
\begin{equation}
    \alpha_I^{\delta\rho}\nabla_\rho\mathcal{H}_+^{I\lambda}=0.
\end{equation}

Going in the opposite direction, as $\nabla_\mu\alpha_I^{\mu\nu}=0$, the field equation can be rewritten as $ \nabla_\mu(\alpha_I^{\mu\nu}\mathcal{H}_+^{I\lambda})=0$. 
Since $\alpha_I^{\mu\nu}\mathcal{H}_+^{I\lambda}$ is a self-dual tensor this implies that the tensor, $F_+^{\mu\nu\lambda}\equiv\alpha^{\mu\nu}_I\mathcal{H}_+^{I\lambda}$ is antisymmetric in $\mu$ and $\nu$. For each $\lambda$, it fulfills $\nabla_\mu F_+^{\mu\nu\lambda}$ = 0, which resembles the equation $\mathrm{d} F_+=0$ of electromagnetism. Then, there is a symmetric tensor $h^+_{\mu\nu}$ such that $F_+^{\mu\nu\lambda}=2 \nabla_{[\mu}h_{\nu]}^{+\,\lambda}$ with the property $\bar\alpha_I^{\mu\nu}\nabla_{[\mu}h_{\nu]}^{+\,\lambda}=0$ (see {\it Lemma 2} of Ref. \cite{Toth:2021dut} for details). Now, we note that $\alpha^{\mu\nu}_I\mathcal{H}_+^{I\lambda}=2 \nabla_{[\mu}h_{\nu]}^{+\,\lambda}$, and multiplying both sides by $\epsilon^{I\mu\nu}$ and using \eqref{Eq:epsilonIab}, we get
\begin{equation}
    \mathcal{H}_+^{I\lambda}=\frac{1}{2}\alpha^{\mu\nu I}\nabla_{[\mu}h_{\nu]}^{+\,\lambda}=i \epsilon^{I\mu\nu}\nabla_\mu h_\nu^{+\,\lambda}\,,
\end{equation}
confirming the relation between fields and potentials in Eq.\eqref{Eq:relationfieldspotentials}.


\section{Spinorial formulation}\label{Appendix_C}

In this appendix we will introduce the spinorial formulation for the variables ${h}^\pm_{\mu\nu}$ and  $\mathcal{H}^{I\mu}_\pm$ in expression \eqref{eq:Hic2hab}. Let us first define the spinor
\begin{equation}
	\Psi=\left(\begin{array}{l}
		h^{+}_{\alpha \beta} \\
		\mathcal{H}_{+}^{I\beta} \\
		h_{-}^{\alpha \beta} \\
		\mathcal{H}^{-}_{\dot I\beta}
	\end{array}\right).
\end{equation}
It is not a usual spinor in the sense that its elements are either rank 2 complex tensorial objects or mixed quantities (containing indexes $I,\dot I, J,\dot J, \ldots $). We should note that the components are related. Namely, the third and fourth elements of the spinor are the complex conjugated of the first and second, respectively, and viceversa. Hence, we can interpret it as a Majorana spinor. 

Let us now denote by $\bf X$ the complex vector space of all $\Psi$, we define now the linear  map $\beta^{\mu}: \bf X \longrightarrow \bf X$ by
\begin{equation}
\beta^{\mu}\Psi=i \left( {\begin{array}{c}
{\bar\alpha^{\dot K \mu}}_{\;\;\;\;\alpha}H^-_{\dot K \beta} \\ - \alpha^{I \mu}_{\;\;\;\; \lambda} h_-^{\lambda\beta} \\  \alpha^{\mu\alpha}_{ K} H_+^{K\beta}  \\  -\bar \alpha^{\mu\lambda}_{\dot I} h^+_{\lambda\beta}  
  \end{array} } \right) \, .
\end{equation}
Then, it seems natural to adopt for this application a matrix-like notation as in Eq. \eqref{eq:psibetamu}. 
\begin{equation}
    \beta^\mu = i\left(\begin{array}{cccc}
		0 & 0 & 0 & {\bar\alpha^{\dot I \mu}}_{\;\;\;\;\alpha} \\
		0 & 0 & - \alpha^{I \mu}_{\;\;\;\; \alpha} & 0 \\
		0 & \alpha^{\mu\alpha}_{ I} & 0 & 0 \\
		-\bar \alpha^{\mu\alpha}_{\dot I} & 0 & 0 & 0
	\end{array}\right),
\end{equation}
keeping in mind that it is convenient to relabel dummy indexes at the end of any operation (for instance we use the first letters of alphabets for free indexes and middle letters in the alphabet for dummy indexes while we reserve the last Greek indexes $\mu,\,\nu,\ldots$ for contraction with the connection $\nabla$).  
We can now define the product of two $\beta^{\mu}$ as the composite operation, $\beta^{\mu}\beta^{\nu}: {\bf X} \longrightarrow {\bf X}$, defined by $(\beta^{\mu}\beta^{\nu})\Psi=\beta^{\mu}(\beta^{\nu}\Psi)$. This is  linear, and leads to
\begin{equation}
\beta^{\mu}(\beta^{\nu}\Psi)=\left(\begin{array}{l}
		\bar\alpha^{\dot K\mu}_{\;\;\;\; \alpha} \bar\alpha^{\nu \lambda}_{\dot K}h^{+}_{\lambda \beta} \\
		\alpha^{I\mu}_{\;\;\;\; \lambda} \alpha^{\nu \lambda}_{  K}\mathcal{H}_{+}^{K\beta} \\
		\alpha^{\mu \alpha}_{K } \alpha^{K\nu}_{\;\;\;\; \lambda}h_{-}^{ \lambda\beta} \\
		\bar \alpha^{\mu \lambda}_{ \dot I} \bar \alpha^{\dot K\nu}_{\;\;\;\;\lambda}\mathcal{H}^{-}_{\dot K\beta}
	\end{array}\right) \, ,
\end{equation}
or in matrix notation
\begin{equation}
    \beta^{\mu}\beta^{\nu}= \,     \left( {\begin{array}{cccc}
   \bar\alpha^{\dot K\mu}_{\;\;\;\; \alpha} \bar\alpha^{\nu \alpha'}_{\dot K}& 0 & 0  & 0  \\
  0  & \alpha^{I\mu}_{\;\;\;\; \lambda} \alpha^{\nu \lambda}_{  I'} & 0 &0 \\
 0 & 0 & \alpha^{\mu \alpha}_{K } \alpha^{K\nu}_{\;\;\;\; \alpha'}  & 0 \\
0 & 0  & 0  &\bar \alpha^{\mu \lambda}_{ \dot I} \bar \alpha^{\dot I'\nu }_{\;\;\;\; \lambda} \\
  \end{array} } \right)\, ,
\end{equation}
where primed indexes are contracted with the same (primed) indexes of the spinor components $\Psi$.

We can use properties in Eqs. \eqref{prop2} and \eqref{prop4} of the $\alpha$ matrices in order to obtain the symmetric and anti-symmetric parts in $\mu$ and $\nu$ of this operation. In particular, one can see that the symmetric part is just
\begin{equation}
\beta^{\mu}(\beta^{\nu}\Psi)+\beta^{\nu}(\beta^{\mu}\Psi)=2g^{\mu\nu} \Psi\, ,
\end{equation}
or in matrix notation
\begin{equation}
    \beta^{\mu}\beta^{\nu}+\beta^{\mu}\beta^{\nu}= 2 g^{\mu\nu}\,     \left( {\begin{array}{cccc}
  1& 0 & 0  & 0  \\
  0  & 1 & 0 &0 \\
 0 & 0 & 1  & 0 \\
0 & 0  & 0  &1 \\
  \end{array} } \right)\, .
\end{equation}
The antisymmetric part will be
\begin{equation}\label{eq:betamunu}
\beta^{\mu}(\beta^{\nu}\Psi)-\beta^{\nu}(\beta^{\mu}\Psi)=2 \left(\begin{array}{l}
		4 \, ^-P^{\mu\nu\;\lambda}_{\;\;\;\;\alpha}h^{+}_{\lambda \beta} \\
		(-1) \, ^+M_{\;\;K}^{I\;\; \mu\nu}\mathcal{H}_{+}^{K\beta} \\
		4 \, ^+P^{\mu\nu\alpha}_{\;\;\;\;\;\;\;\lambda}h_{-}^{\lambda\beta } \\
		(-1) \, ^-M_{\dot I}^{\;\;\dot K \mu\nu}\mathcal{H}^{-}_{\dot K\beta}
	\end{array}\right)\, ,
\end{equation}
which in matrix notation amounts to
\begin{equation}
    \beta^{\mu}\beta^{\nu}-\beta^{\mu}\beta^{\nu}= 2 \,     \left( {\begin{array}{cccc}
  4 \, ^-P^{\mu\nu\;\alpha'}_{\;\;\;\;\alpha}& 0 & 0  & 0  \\
  0  & (-1) \, ^+M_{\;\;I'}^{I\;\; \mu\nu} & 0 &0 \\
 0 & 0 &  4 \, ^+P^{\mu\nu\alpha}_{\;\;\;\;\;\;\;\alpha'}  & 0 \\
0 & 0  & 0  &(-1) \, ^-M_{\dot I}^{\;\;\dot I' \mu\nu} \\
  \end{array} } \right)\, .
\end{equation}
We can now define the composite operation $\beta_5$, the so called ``chiral" matrix, as the linear map $\beta_5: {\bf X} \longrightarrow {\bf X}$  by $\beta_5   \equiv    \frac{i}{4!}\epsilon_{\mu\nu\rho\delta} \beta^{\mu} \beta^{\nu} \beta^{\rho} \beta^{\delta}$, or equivalently $\beta_5  =  \frac{i}{4!}\epsilon_{\mu\nu\rho\delta} \beta^{[\mu} \beta^{\nu]} \beta^{[\rho} \beta^{\delta]} $. One can then see that 
\begin{equation}
\beta_5\Psi=\frac{i}{6} \epsilon_{\mu\nu\rho\delta}\left(\begin{array}{l}
		4 \left[^+P^{\mu\nu}\right]_{\alpha}^{\ \gamma}  \left[^+P^{\rho\delta}\right]_{\gamma}^{\ \lambda}h^{+}_{\lambda\beta} \\
		\frac{1}{4}\left[^+M^{\mu\nu}\right]^{I}_{\ K}  \left[^+M^{\rho\sigma}\right]^{K}_{\hspace{0.15cm}L}\mathcal{H}_{+}^{L\beta} \\
		4 \left[^-P^{\mu\nu}\right]^{\alpha}_{\ \gamma}  \left[^-P^{\rho\sigma}\right]^{\gamma}_{\hspace{0.15cm}\lambda}h_{-}^{\lambda\beta } \\
		\frac{1}{4} \left[^-M^{\mu\nu}\right]_{\dot I}^{\ \dot K}  \left[^-M^{\alpha\beta}\right]_{\dot K}^{\hspace{0.15cm}\dot J}\mathcal{H}^{-}_{\dot J\beta}
	\end{array}\right)= \left(\begin{array}{l}
		-h^{+}_{\alpha \beta} \\
		-\mathcal{H}_{+}^{I\beta} \\
		h_{-}^{\alpha \beta} \\
		\mathcal{H}^{-}_{\dot I\beta}
	\end{array}\right)\, ,
\end{equation}
or in matrix notation
\begin{eqnarray}
\beta_5 & = & \left( {\begin{array}{cccc}
- 1 & 0 & 0  &0 \\
  0  & -1 & 0 &0 \\
 0 & 0  & 1   & 0 \\
 0  & 0  & 0  &1 \\
  \end{array} } \right) \, .
  \end{eqnarray}
  The map $\beta_5$ has the following properties
\begin{eqnarray}
\beta_5^2=   \left( {\begin{array}{cccc}
 1 & 0 & 0  &0 \\
  0  & 1 & 0 &0 \\
 0 & 0  & 1   & 0 \\
 0  & 0  & 0  &1\\
  \end{array} } \right)   \, , \hspace{.5cm} \{\beta_5, \beta^{\mu}\} = 0 \, .
\end{eqnarray}
It is worth noting that a duality transformation can be implemented by means of the linear operation $T_{\theta}: {\bf X} \longrightarrow {\bf X}$, with $T_{\theta}=e^{i\theta \beta_5}$, $\theta\in \mathbb R$. Let ${\bf X}^\dagger$ be now the dual space, namely, the space of linear functionals over ${\bf X}$. Given $\Psi\in {\bf X}$, we then define $\bar \Psi\in {\bf X}^\dagger$ by 
\begin{equation}
\bar{\Psi}=\left(h^{+}_{\alpha \beta},\mathcal{H}_{+}^{I\beta},h^{-}_{\alpha \beta},
	\mathcal{H}_{-}^{I\beta}\right)\, . \label{eq:barpsi}
\end{equation}
We can now define an inner product. Although the product $\bar \Psi \Psi \in \mathbb C$ is well defined, it does not produce a positive real number. We can define a (positive-definite) inner product as follows
\begin{equation}
\label{eq:product} \langle \Psi_1,\Psi_2 \rangle =\frac{1}{s}\, \int d^4x \sqrt{-g} \, \bar \Psi_1 \,\delta \,  \Psi_2 \, ,
\end{equation} 
where $s$ is an arbitrary positive real constant with dimensions of action, and $\delta: {\bf X} \longrightarrow {\bf X}^\dagger$ is a linear application between ${\bf X}$ and its dual space ${\bf X}^\dagger$ defined by
\begin{eqnarray}
\delta\Psi=\left(\begin{array}{l}
        h_{-}^{\alpha \beta} \\
		\mathcal{H}^{-}_{\dot I\beta}\\
		h^{+}_{\alpha \beta} \\
		\mathcal{H}_{+}^{I\beta}
	\end{array}\right),  \label{eq:delta} \qquad {\rm with} \qquad
\delta = \left( {\begin{array}{cccc}
 0& 0 & 1  &0 \\
  0  & 0 & 0 & \gamma^I_{\dot I} \\
 1  & 0  & 0   & 0 \\
 0  & \gamma^{ I}_{\dot I}  & 0  &0 \\
  \end{array} } \right)\, ,
\end{eqnarray}
(where $\gamma^I_{\dot I}$ is a mixed Kronecker delta defined in Eq. \eqref{eq:gammas}).  \footnote{This product is similar to the one of a Dirac field $\langle \Psi_1,\Psi_2 \rangle =\frac{1}{s}\, \int d^4x \sqrt{-g} \, \Psi^{\dagger}_1  \,  \Psi_2$ but with matrix $\delta$ playing the role of $\gamma^0$ such that $ \Psi^{\dagger} \equiv \bar \Psi \delta$ (see e.g. \cite{Fujikawa:1980eg}).} This operation yields $\bar \Psi \delta \Psi = 2 h_{-}^{\alpha \beta}h^{+}_{\alpha \beta}+2\delta^{\dot I}_I\mathcal{H}^{-}_{\dot I\beta}\mathcal{H}_{+}^{I\beta}\geq0$ since it involves products of complex numbers times their complex conjugate.  Morover, one can see that $\left< \Psi_1, \Psi_2\right>=\left< \Psi_2, \Psi_1\right> $. In addition, one can trivially check linearity of this inner product with respect to the second variable. Let us note that this inner product allows us to define a basis in ${\bf X}$, and hence, what is the meaning of the trace of linear operators like $\beta^{\mu}$ and their products. 

From the algebraic perspective, a choice of basis is given by
\begin{equation}
	\psi_1=\frac{\sqrt{s}}{4\sqrt{2}}\left(\!\!\begin{array}{l}
		g_{\alpha \nu}g_{\beta\lambda} \\
		0 \\
		g^{\alpha \nu}g^{\beta\lambda} \\
		0
	\end{array}\!\!\!\right)\!,\, \psi_2=\frac{\sqrt{s}}{4\sqrt{2}}\left(\!\!\begin{array}{l}
		0 \\
		\eta^{IJ}g^{\beta\lambda} \\
		0 \\
		\eta_{\dot I\dot J}g_{\beta\lambda}
	\end{array}\!\!\right)\!,\, \psi_3=\frac{i\sqrt{s}}{4\sqrt{2}}\left(\!\!\begin{array}{l}
		g_{\alpha \nu}g_{\beta\lambda} \\
		0 \\
		-g^{\alpha \nu}g^{\beta\lambda} \\
		0
	\end{array}\!\!\!\right)\!,\, \psi_4=\frac{i\sqrt{s}}{4\sqrt{2}}\left(\!\!\begin{array}{l}
		0 \\
		\eta^{IJ}g^{\beta\lambda} \\
		0 \\
		-\eta_{\dot I\dot J}g_{\beta\lambda}
	\end{array}\!\!\right)\!,
\end{equation}
where we include $\sqrt{s}$ in the normalization as given by the inner product. One can directly check that $\bar \psi_n \delta \psi_m=s\,\delta_{nm}$ (assuming contraction of indexes $J$ and $\dot J$ with the mixed Kronecker delta $\gamma_J^{\dot J}$ given in Eq. \eqref{eq:gammas}). Hence, the (local) trace of any linear map $F: {\bf X} \longrightarrow {\bf X}$ is given by the usual definition 
\begin{equation}\label{eq:trace-oper}
\operatorname{Tr}[F]=\sum_n \bar \psi_n (\delta( F \psi_n)) .
\end{equation}
For instance, one can see that
\begin{equation}\label{eq:trace-oper}
\frac{1}{s}\operatorname{Tr}[\beta^{\mu}\beta^{\nu}]= 4g_{\mu\nu},\quad \frac{1}{s}\operatorname{Tr}[\beta^{\mu}\beta^{\nu}\beta^{\rho}\beta^{\sigma}]= 4 g_{\mu\nu}g_{\rho\sigma}-4g_{\mu\rho}g_{\nu\sigma}+4g_{\mu\sigma}g_{\rho\nu},
\end{equation}
as expected, as well as, 
\begin{equation}\label{eq:trace-oper}
\frac{1}{s}\operatorname{Tr}[\beta_5\beta^{\mu}\beta^{\nu}]= 0,\quad \frac{1}{s}\operatorname{Tr}[\beta_5\beta^{\mu}\beta^{\nu}\beta^{\rho}\beta^{\sigma}]= 4i\epsilon^{\mu\nu\rho\sigma}.
\end{equation}
We must note that the normalization of these (local) traces involves the normalization (by the constant $s$) of the inner product in Eq. \eqref{eq:product}. However, it cancels upon integration. In our case, the Dirac action of gravitational perturbations is normalized to $s=4$ in the units adopted in this manuscript. 

\section{Curvature operators acting on spinors and their traces}\label{Appendix_D}

In this Appendix we will start computing the action of the operator $W_{\rho \mu}=\left[\nabla_\rho, \nabla_\mu\right]=(\nabla_\rho \nabla_\mu- \nabla_\mu\nabla_\rho)$ on the components of $\Psi$. Let us start with an arbitrary vector $A^\nu$. One can easily see that
\begin{equation}
\left[\nabla_\rho, \nabla_\mu\right] A_{\nu} = R_{\rho \mu \nu}{}^{\sigma}A_{\sigma}= R_{\rho \mu \sigma\delta}\;(\Sigma^{\sigma\delta})_{\nu}{}^{\gamma}\;A_{\gamma},
\end{equation}
where
\begin{equation}
(\Sigma^{\sigma\delta})_{\nu}{}^{\gamma}=g^{\gamma[\sigma }\delta^{\delta] }_{\nu}   ,
\end{equation}
is the (well-known) generator of the $(1/2,1/2)$ (real) representation of the Lorentz group. 

Let us consider the potential $h^+_{\nu\lambda}$:
\begin{equation}\label{eq:wmunutohpm}
\left[\nabla_\rho, \nabla_\mu\right] h^\pm_{\nu\lambda} = R_{\rho \mu \nu}{}^{\sigma}h^\pm_{\sigma\lambda}+ R_{\rho \mu \lambda}{}^{\sigma}h^\pm_{\nu\sigma} = R_{\rho \mu \sigma\delta}\;(\Sigma^{\sigma\delta})_{\nu\lambda}{}^{\gamma\epsilon}\;h^\pm_{\gamma\epsilon},
\end{equation}
where we define
\begin{equation}\label{eq:sigmahnl}
(\Sigma^{\sigma \delta}){}_{\nu\lambda}{}^{\gamma\epsilon}=4\delta^{[\sigma|}_{(\nu}
\delta^{(\epsilon}_{\lambda)}g^{\gamma)|\delta]},
\end{equation}
which is symmetric in $(\nu\lambda)$ and $(\gamma,\epsilon)$ and antisymmetric in $(\sigma,\delta)$. Actually, it is the generator of the $(1,1)$ (real) representations of the Lorentz group (the one corresponding to 2-rank symmetric and traceless tensors).

In a similar way, we can determine how the operator $W_{\rho \mu}$ acts on ${\cal H}^{I\lambda}$. In order to proceed, let us note first that for the Fierz tensor in the TT gauge (see Appendix \ref{Appendix_E})
\begin{equation}
\left[\nabla_\rho, \nabla_\mu\right] {}^{\pm}{\rm F}^{\sigma\delta\lambda} = - R_{\rho \mu \epsilon}{}^{\sigma}{}^{\pm}{\rm F}^{\epsilon\delta\lambda}-R_{\rho \mu \epsilon}{}^{\delta}{}^{\pm}{\rm F}^{\sigma\epsilon\lambda}- R_{\rho \mu \epsilon}{}^{\lambda}{}^{\pm}{\rm F}^{\sigma\delta\epsilon}= R_{\rho \mu \sigma\delta}\;(\Sigma^{\sigma \delta}){}^{\nu\lambda\epsilon}{}_{\alpha\beta\gamma}\;{}^{\pm}{\rm F}^{\alpha\beta\gamma},
\end{equation}
where
\begin{equation}
(\Sigma^{\sigma \delta}){}^{\nu\lambda\epsilon}{}_{\alpha\beta\gamma}=
4\delta^{[\nu  }_{[\alpha} \delta^{\lambda]  }_{\beta]  }\delta^{[\delta|  }_{\gamma }g^{|\sigma]  \epsilon  }
+\delta^{\lambda  }_{[\alpha} \delta^{[\sigma| }_{\beta]  }\delta^{\epsilon  }_{\gamma }  g^{|\delta]  \nu  } 
+ \delta^{\delta  }_{[\alpha} \delta^{[\lambda |}_{\beta]  }\delta^{\epsilon  }_{\gamma }  g^{\sigma  |\nu]  }
-\delta^{\sigma  }_{[\alpha} \delta^{[\lambda |}_{\beta]  }\delta^{\epsilon  }_{\gamma }  g^{\delta  |\nu]  }
+ \delta_{[\alpha}^{[\lambda|  } \delta^{\sigma  }_{\beta]  }\delta^{\epsilon  }_{\gamma } g^{\delta  |\nu]  }
-\delta_{[\alpha}^{[\lambda|  } \delta^{\delta  }_{\beta]  }\delta^{\epsilon  }_{\gamma } g^{\sigma  |\nu]  }
\end{equation}
Now, out of the relation between ${\cal H}_+^{I\lambda}$ and ${}^{+}{\rm F}^{\mu\nu\lambda}$ in Eq. \eqref{Eq:fierz_fields}, and noting that here $I=1,2,3,$ we can easily define the corresponding linear map for $W_{\rho\mu}{\cal H}_+^{I\lambda}$ as
\begin{equation}\label{eq:wmunutoHp}
\left[\nabla_\rho, \nabla_\mu\right]{\cal H}_+^{I\lambda} = R_{\rho\mu\sigma\delta} {}^{+}(\Sigma^{\sigma \delta}){}^{I\lambda}{}_J{}_{\epsilon}{\cal H}_+^{J\epsilon},
\end{equation}
where
\begin{equation}\label{eq:sigmapHil}
^{+}(\Sigma^{\sigma \delta}){}^{I\epsilon}{}_J{}_{\gamma}=\frac{1}{4}(\Sigma^{\sigma \delta}){}^{\nu\lambda\epsilon}{}_{\alpha\beta\gamma}h^I_{I'}h^{J'}_{J}\alpha^{I'}{}_{\nu\lambda}\alpha_{J'}^{\alpha\beta}= 
-\frac{1}{2}i \delta^{\epsilon}_{\gamma} \epsilon^I{}_{JK} \alpha^{K\sigma\delta},
\end{equation}
which amounts to the generator of the $(0,1)\bigotimes(1/2,1/2)$ representation of the Lorentz group. On the other hand, for ${\cal H}_-^{I\lambda}$ we have a similar expression 
\begin{equation}\label{eq:wmunutoHm}
\left[\nabla_\rho, \nabla_\mu\right]{\cal H}_-^{I\lambda} = R_{\rho\mu\sigma\delta} {}^{-}(\Sigma^{\sigma \delta}){}^{I\lambda}{}_J{}_{\epsilon}{\cal H}_-^{J\epsilon},
\end{equation}
but with
\begin{equation}\label{eq:sigmamHil}
^{-}(\Sigma^{\sigma \delta}){}^{\dot I\epsilon}{}_{\dot J}{}_{\gamma}=\frac{1}{4}(\Sigma^{\sigma \delta}){}^{\nu\lambda\epsilon}{}_{\alpha\beta\gamma}\bar\alpha^{\dot I}{}_{\nu\lambda}\bar\alpha_{\dot J}^{\alpha\beta}= 
\frac{1}{2}i \delta^{\epsilon}_{\gamma} \epsilon^{\dot I}{}_{\dot J\dot K} \bar\alpha^{\dot K\sigma\delta},
\end{equation}
the generator of the $(1,0)\bigotimes(1/2,1/2)$ representation of the Lorentz group. Taking into account Eqs. \eqref{eq:wmunutohpm}, \eqref{eq:wmunutoHp} and \eqref{eq:wmunutoHm}, we obtain $W_{\rho \mu}\Psi$ as given in Eq. \eqref{Eq:defWmunu}. 

With this, and the results of the previous appendix, one can compute the traces of several operators acting on the spinors $\Psi$. Let us start with the operator $\mathcal{Q}$ defined in Eq. \eqref{Eq:defQ}. We are interested in the trace of $\beta_5\mathcal{Q}$. After some lengthy but simple calculations, one can see that 
  \begin{equation}
  	{\rm Tr}[\beta_5\mathcal{Q}]=-i\frac{1}{4}\epsilon_{\mu\nu\rho\sigma}R^{\mu\nu\rho\sigma},
  \end{equation}
which equals zero due to the first Bianchi identity of the Riemann tensor. 

The next operator we are interested in is $W_{\mu\nu}W^{\mu\nu}$. From the definition in Eq. \eqref{Eq:defWmunu}, one can see that
\begin{equation}\label{Eq:defWmunu2}
W^{\mu \nu}W_{\mu \nu} \Psi = R_{\mu \nu \sigma \delta}R^{\mu \nu}{}_{\sigma' \delta'}\left(\begin{array}{l}
		(\Sigma^{\sigma' \delta'}){}_{\alpha \beta}{}^{\gamma'\epsilon'}(\Sigma^{\sigma \delta}){}_{\gamma'\epsilon'}{}^{\gamma\epsilon}h^{+}_{\gamma \epsilon} \\
		^{+}(\Sigma^{\sigma' \delta'}){}^{I\alpha}{}_{J'}{}_{\gamma'}{}^{+}(\Sigma^{\sigma \delta}){}^{J'\gamma'}{}_J{}_{\gamma}\mathcal{H}_{+}^{J\gamma} \\
		(\Sigma^{\sigma' \delta'}){}^{\alpha \beta}{}_{\gamma'\epsilon'}(\Sigma^{\sigma \delta}){}^{\gamma'\epsilon'}{}_{\gamma\epsilon}h_{-}^{\gamma \epsilon} \\
		{}^{-}(\Sigma^{\sigma' \delta'}){}_{\dot I\alpha}{}{}^{\dot J'\gamma'}{}^{-}(\Sigma^{\sigma \delta}){}_{\dot J'\gamma'}{}{}^{\dot J\gamma}\mathcal{H}^{-}_{\dot J\gamma}
	\end{array}\right).
\end{equation}
The trace of $\beta_5 W^{\mu \nu}W_{\mu \nu}$ is given by \footnote{We have computed all these tensorial calculations with a notebook \cite{jolmedo2025} based on the xTensor package of Mathematica \cite{MartinGarcia:2008}.}
  \begin{equation}
  	{\rm Tr}[\beta_5W^{\mu \nu}W_{\mu \nu}]=i\epsilon_{\mu\nu\rho\sigma}R^{\mu\nu\gamma\delta}R^{\rho\sigma}{}_{\gamma\delta}.
  \end{equation}
Finally, we also need to compute the expression of the operator $\mathcal{Q}^2$. Given the definition of in Eq. \eqref{Eq:defQ}, one can see that 
  \begin{equation}
  	\mathcal{Q}^2 \Psi \equiv \frac{1}{4} R_{\mu \nu \sigma \delta}R_{\mu' \nu' \sigma' \delta'}\left(\begin{array}{l}
		16 \, ^-P^{\mu'\nu'\;\;\;\lambda'}_{\;\;\;\;\;\;\;\alpha}(\Sigma^{\sigma' \delta'}){}_{\lambda'\beta}{}^{\gamma'\epsilon'} \, ^-P^{\mu\nu\;\;\;\lambda}_{\;\;\;\;\;\gamma'}(\Sigma^{\sigma \delta}){}_{\lambda\delta'}{}^{\gamma\epsilon}h^{+}_{\gamma \epsilon} \\
		 {}^+M_{\;\;J'}^{I\;\; \mu'\nu'}{}^{+}(\Sigma^{\sigma' \delta'}){}^{J'\alpha}{}_{K'}{}_{\gamma'}\,{}^+M_{\;\;J}^{K'\;\; \mu\nu}{}^{+}(\Sigma^{\sigma \delta}){}^{J\gamma'}{}_K{}_{\gamma}\mathcal{H}_{+}^{K\gamma} \\
		16 \, ^+P^{\mu'\nu'\alpha}_{\;\;\;\;\;\;\;\;\;\lambda'}(\Sigma^{\sigma' \delta'}){}^{\lambda'\beta}{}_{\gamma'\epsilon'}\, ^+P^{\mu\nu\gamma'}_{\;\;\;\;\;\;\;\lambda}(\Sigma^{\sigma \delta}){}^{\lambda\delta'}{}_{\gamma\epsilon}h_{-}^{\gamma \epsilon} \\
		{}^-M_{\dot I}^{\;\;\dot J' \mu'\nu'}{}^{-}(\Sigma^{\sigma' \delta'}){}_{\dot J'\alpha}{}{}^{\dot K'\gamma'}\,{}^-M_{\dot K'}^{\;\;\dot J \mu\nu}{}^{-}(\Sigma^{\sigma \delta}){}_{\dot J\gamma'}{}{}^{\dot K\gamma}\mathcal{H}^{-}_{\dot K\gamma}
	\end{array}\right).\label{Eq:defQ2}
  \end{equation}
Again, after some lengthy but simple calculations (see notebook \cite{jolmedo2025}), we obtain for the trace of $\beta_5\mathcal{Q}^2$
  \begin{equation}
  	{\rm Tr}[\beta_5\mathcal{Q}^2]=-i\frac{1}{8}\left(10\epsilon_{\mu\nu\rho\sigma}R^{\mu}{}_{\mu'}R^{\mu'\nu\rho\sigma}+5R\epsilon_{\mu\nu\rho\sigma}R^{\mu\nu\rho\sigma}+6\epsilon_{\mu\nu\rho\sigma}R^{\mu\nu\gamma\delta}R^{\rho\sigma}{}_{\gamma\delta}+2\epsilon_{\delta\nu\rho\sigma}R^{\mu\nu\gamma\delta}R^{\rho\sigma}{}_{\gamma\mu}\right).
  \end{equation}
Using the first Bianchi identity for the Riemann tensor, one can see that the first and second addends are identically zero while the last addend is proportional to the third one. The final result is 
\begin{equation}
  	{\rm Tr}[\beta_5\mathcal{Q}^2]=-i\frac{5}{8}\epsilon_{\mu\nu\rho\sigma}R^{\mu\nu\gamma\delta}R^{\rho\sigma}{}_{\gamma\delta}.
  \end{equation}
  

\section{Einstein-Hilbert vs Maxwellian gravity  in curved space}\label{Appendix_E}

In our computations to obtain a Maxwell-like description, we begin by working in the TT gauge and then adding a total derivative to the standard linearized gravity Lagrangian. On a flat background, since a total derivative is defined with a partial derivative, these operations commute and no problem arises. However, when generalizing the background metric the situation changes: the operations of taking a total derivative and imposing the TT gauge no longer necessarily commute. To make it explicit, we will first study a metric perturbation on an originally curved background. Then, we will go back to the formulation with the Maxwell-like Lagrangian in Minkowski and in the TT gauge, Eq. \eqref{Eq:GWMaxwelllikeL}, and generalize this action to a curved background in order to compare the equations of motion. 

The perturbed Einstein-Hilbert action in a generic background $g_{\mu\nu}$ can be written as \cite{Andersson:2020gsj},
\begin{flalign}
	S_{l i n}\left(h_{\mu \nu}\right)= & \int \mathrm{d}^4 x \sqrt{g}\left(\frac{1}{2} \nabla_\gamma h \nabla^{\gamma} h-\frac{1}{2} \nabla^\gamma h^{\mu\nu} \nabla_\gamma h_{\mu\nu}\right.  \left.-\nabla_\mu h \nabla_\nu h^{\mu\nu}+\nabla_\mu h_{\gamma\nu} \nabla^\gamma h^{\mu\nu}\right)\nonumber\\
	= & \int \mathrm{d}^4 x \sqrt{g}h^{\mu\nu}	\hat{D}_{\mu\nu}{ }^{\gamma \delta} h_{\gamma \delta}\,, \label{eq:Eq:lin-HE}
\end{flalign}
where the second line was obtained through integration by parts and $\hat{D}_{\alpha\beta}$ is given by
\begin{equation}
	\begin{aligned}
		\hat{D}_{\mu \nu}{ }^{\gamma \delta}=  \frac{1}{2}\left(\delta_\mu^\gamma \delta_\nu^\delta \nabla_\mu \nabla^\mu-g_{\mu\nu} g^{\gamma \delta} \nabla_\alpha \nabla^\alpha+g^\gamma \nabla_\mu \nabla_\nu\right. \left.+g_{\mu\nu} \nabla^r \nabla^\delta-\delta_\nu^\delta \nabla^\gamma \nabla_\mu-\delta_\mu^\delta \nabla^\gamma \nabla_\nu\right) .
	\end{aligned}
\end{equation}
In this way the Euler-Lagrange equations are easily expressed by 
\begin{equation}
	\hat{D}_{\mu\nu}{ }^{\gamma \delta} h_{\gamma\delta }=0\,.
\end{equation}
Next, considering for convenience the trace reverse tensor $\bar{h}^{\mu\nu}=h^{\mu\nu}-\frac{1}{2} h g^{\mu\nu}$, we rewrite the equations of motion as 
\begin{equation}
	\nabla_\lambda \nabla^\lambda \bar{h}_{\mu \nu}+\nabla_\lambda \nabla_\delta \bar{h}^{\lambda \delta} g_{\mu \nu} 
	-\nabla^\lambda \nabla_\mu \bar{h}_{\lambda \nu}-\nabla^\lambda \nabla_\nu \bar{h}_{\lambda \mu}=0 \,.\label{Eq:eomTracereverse}
\end{equation}
Taking the Lorenz gauge
\begin{equation}
	\nabla_\mu \bar{h}^{\mu\nu }=\nabla_\mu\left(h^{\mu\nu}-\frac{1}{2} h g^{\mu\nu}\right)=0 \,,
\end{equation}
we find
\begin{equation}
	\nabla_\lambda \nabla^\lambda \bar{h}_{\mu \nu}
	-\nabla^\lambda \nabla_\mu \bar{h}_{\lambda \nu}-\nabla^\lambda \nabla_\nu \bar{h}_{\lambda \mu}=0\,.
\end{equation}
Moreover, from the trace of Eq.\eqref{Eq:eomTracereverse} in the Lorenz gauge one gets
\begin{equation}
	2 \nabla^\lambda \nabla^\mu \bar{h}_{\lambda \mu}=\nabla_\lambda \nabla^\lambda h=0\, ,
\end{equation}
such that the trace of $h$ decouples, since $h$ is not explicitly coupled to $\bar{h}_{\mu\nu}$. Considering a theory without sources, the Ricci curvature of $g_{\mu\nu}$  will vanish and we may finally write the evolution of the perturbations as 
\begin{equation}
	\nabla^\lambda \nabla_\mu {h}_{\lambda \nu}+\nabla^\lambda \nabla_\nu {h}_{\lambda \mu}-\nabla_\lambda \nabla^\lambda {h}_{\mu \nu}=0\,,
\end{equation}
or
\begin{equation}
	\nabla^\lambda \nabla_\lambda h_{\mu \nu}-2 R_{\nu \lambda \sigma \mu} h^{\lambda \sigma}=0\,,
\end{equation}
by introducing the Riemann tensor.

Let us now recover the description from the previous section \ref{sectionflatGWs}. On a flat background the similitude to Maxwell theory is made more evident when working with the Fierz tensor \cite{Novello:2002cj,Toth:2021dut}\footnote{The definition of the Fierz tensor given in Eq. \eqref{eq:fiert} agrees with Ref. \cite{Novello:2002cj} but it does not coincide with the one in \cite{Toth:2021dut}. }
\begin{equation}\label{eq:fiert}
	F_{a b c}=\frac{1}{2}\left(\partial_b h_{a c}-\partial_a h_{b c}+\partial_d h^d{ }_b \eta_{a c}-\partial_d h^d{ }_a \eta_{b c}+\partial_a h \, \eta_{b c}-\partial_b h \, \eta_{a c}\right),
\end{equation}
that satisfies the conditions
\begin{flalign}
	& F_{a b c}+F_{b a c}=0\,, \\
	& F_{a b c}+F_{b c a}+F_{c a b}=0 .
\end{flalign}
We also define its Hodge dual as $\tilde F_{a b c}=\frac{1}{2}{\epsilon_{ab}}^{de} F_{d e c}$. Hence, the self- and anti-self-dual Fierz tensors are
\begin{equation}\label{eq:fiert}
	{}^{\pm}F^{a b c}=\frac{1}{\sqrt{2}}(F^{a b c}\pm i \tilde F^{a b c}).
\end{equation}
We can also relate these self- and anti-self-dual Fierz tensors with the new variables field variables $\mathcal{H}^{cI}_\pm$ as
\begin{equation}
{}^{+}F^{abc}= -\frac{1}{2}\alpha^{ab}_I \mathcal{H}^{cI}_+,\qquad {}^{-}F^{abc}=-\frac{1}{2}\bar{\alpha}^{ab}_I \mathcal{H}^{cI}_-\,.\label{Eq:fierz_fields}
\end{equation}

In the TT gauge, the Fierz tensor takes the form ${\rm 
F}_{a b c}=\frac{1}{2}\left(\partial_b h_{a c}-\partial_a h_{b c}\right)$. The  Maxwell-like Lagrangian from Eq.\eqref{Eq:GWMaxwelllikeL}   can be written as 
\begin{equation}
	\mathcal{L}^{TT}_{{FP}}=-\frac{1}{2}{\rm 
F}^{abc} {\rm 
F}_{abc}+\frac{1}{2}\tilde {\rm 
F}^{abc} \tilde {\rm 
F}_{abc}=-\frac{1}{2}{}^{+}{\rm 
F}^{abc} {}^{+}{\rm 
F}_{abc}-\frac{1}{2}{}^{-}{\rm 
F}^{abc} {}^{-}{\rm 
F}_{abc}. \label{Eq:FierzTensorLagrangian}	
\end{equation}
This Lagrangian agrees with Eqs. \eqref{eq:LG2} and \eqref{eq:LG3}.

One can see that 
\begin{equation}
	{\rm 
F}_{0ij} = -\frac{1}{2} e_{ij}, \qquad {\rm 
F}_{ijk} = -\frac{1}{2} \epsilon_{ijl}{b^l}_{k},\label{Eq:FierzTensorLagrangian}	
\end{equation}
with $e_{ij}$ and $b_{ij}$ defined in Eq. \eqref{eq:enb}. In the same way, we have for the Hodge dual 
\begin{equation}
	\tilde {\rm 
F}_{0ij} = - \frac{1}{2} b_{ij}, \qquad \tilde {\rm 
F}_{ijk} = \frac{1}{2} \epsilon_{ijl}{e^l}_{k}.\label{Eq:FierzTensorLagrangian}	
\end{equation}

If we generalize our theory to a curved background from the Lagrangian above we will find the equations fo motion for the metric perturbations to be
\begin{equation}
	\nabla^\lambda \nabla_\mu {h}_{\lambda \nu}+\nabla^\lambda \nabla_\nu {h}_{\lambda \mu}-2\nabla_\lambda \nabla^\lambda {h}_{\mu \nu}=0\, ,\label{Eq:EomFierzcurved}	
\end{equation}
or 
\begin{equation}
	\nabla^\lambda \nabla_\lambda h_{\mu \nu}- R_{\nu \lambda \sigma \mu} h^{\lambda \sigma}=0\,.
\end{equation}
Comparing with the equation of motion obtained in Eq.\eqref{Eq:EomFierzcurved} we find a difference of a factor 2 in the coupling with Riemann curvature between the two equations. 

As a conclusion, the extension of the Maxwell-like Lagrangian developed in section \ref{sectionflatGWs} to a curved background would lead to a different theory than what would be given by General Relativity. The divergence arises from applying the TT gauge conditions before adding a total derivative to rearrange the Lagrangian. 

On the other hand, when considering the propagation of gravitational waves in a curved background in the geometric optics approximation these descriptions coincide. Following~\cite{Maggiore:2007ulw}, we take $h\sim O(|h_{\mu\nu}|)$,   $\lambdabar$ to represent the amplitude wavelength of the gravitational perturbation and $L_B$ as the scale of the spatial variation of the background, such that $\lambdabar \ll L_B$. As discussed in \cite{Maggiore:2007ulw}, on a curved background  $h\ll\lambdabar/L_B\ll1$ and as a consequence $\nabla^\lambda \nabla_\lambda {h}_{\mu \nu}=O(h / \lambdabar^2)$ since ${R}_{\mu \lambda  \sigma\nu} {h}^{\lambda \sigma}=O(h / L_B^2)$. Then, it follows that in the limit $\lambdabar \ll L_B$ both constructions coincide.


\bibliography{AnomalyGravitationalWaves.bib}

\end{document}